\documentclass[aps,prd,twocolumn,notitlepage,
showpacs,superscriptaddress,floatfix]{revtex4-1} 
\usepackage{graphicx} 
\usepackage{dcolumn} 
\usepackage{bm} 
\usepackage{amssymb} 
\usepackage{hyperref} 

\begin{document}

\title{Measurement of inclusive double-differential \texorpdfstring{$\nu_\mu$}{nu} charged-current cross section with improved acceptance in the T2K off-axis near detector}


\newcommand{\INSTHD}{\affiliation{University Autonoma Madrid, Department of Theoretical Physics, 28049 Madrid, Spain}}
\newcommand{\INSTEE}{\affiliation{University of Bern, Albert Einstein Center for Fundamental Physics, Laboratory for High Energy Physics (LHEP), Bern, Switzerland}}
\newcommand{\INSTFE}{\affiliation{Boston University, Department of Physics, Boston, Massachusetts, U.S.A.}}
\newcommand{\INSTD}{\affiliation{University of British Columbia, Department of Physics and Astronomy, Vancouver, British Columbia, Canada}}
\newcommand{\INSTGA}{\affiliation{University of California, Irvine, Department of Physics and Astronomy, Irvine, California, U.S.A.}}
\newcommand{\INSTI}{\affiliation{IRFU, CEA Saclay, Gif-sur-Yvette, France}}
\newcommand{\INSTGB}{\affiliation{University of Colorado at Boulder, Department of Physics, Boulder, Colorado, U.S.A.}}
\newcommand{\INSTFG}{\affiliation{Colorado State University, Department of Physics, Fort Collins, Colorado, U.S.A.}}
\newcommand{\INSTFH}{\affiliation{Duke University, Department of Physics, Durham, North Carolina, U.S.A.}}
\newcommand{\INSTBA}{\affiliation{Ecole Polytechnique, IN2P3-CNRS, Laboratoire Leprince-Ringuet, Palaiseau, France }}
\newcommand{\INSTEG}{\affiliation{University of Geneva, Section de Physique, DPNC, Geneva, Switzerland}}
\newcommand{\INSTDG}{\affiliation{H. Niewodniczanski Institute of Nuclear Physics PAN, Cracow, Poland}}
\newcommand{\INSTCB}{\affiliation{High Energy Accelerator Research Organization (KEK), Tsukuba, Ibaraki, Japan}}
\newcommand{\INSTED}{\affiliation{Institut de Fisica d'Altes Energies (IFAE), The Barcelona Institute of Science and Technology, Campus UAB, Bellaterra (Barcelona) Spain}}
\newcommand{\INSTEC}{\affiliation{IFIC (CSIC \& University of Valencia), Valencia, Spain}}
\newcommand{\INSTEI}{\affiliation{Imperial College London, Department of Physics, London, United Kingdom}}
\newcommand{\INSTGF}{\affiliation{INFN Sezione di Bari and Universit\`a e Politecnico di Bari, Dipartimento Interuniversitario di Fisica, Bari, Italy}}
\newcommand{\INSTBE}{\affiliation{INFN Sezione di Napoli and Universit\`a di Napoli, Dipartimento di Fisica, Napoli, Italy}}
\newcommand{\INSTBF}{\affiliation{INFN Sezione di Padova and Universit\`a di Padova, Dipartimento di Fisica, Padova, Italy}}
\newcommand{\INSTBD}{\affiliation{INFN Sezione di Roma and Universit\`a di Roma ``La Sapienza'', Roma, Italy}}
\newcommand{\INSTEB}{\affiliation{Institute for Nuclear Research of the Russian Academy of Sciences, Moscow, Russia}}
\newcommand{\INSTHA}{\affiliation{Kavli Institute for the Physics and Mathematics of the Universe (WPI), The University of Tokyo Institutes for Advanced Study, University of Tokyo, Kashiwa, Chiba, Japan}}
\newcommand{\INSTCC}{\affiliation{Kobe University, Kobe, Japan}}
\newcommand{\INSTCD}{\affiliation{Kyoto University, Department of Physics, Kyoto, Japan}}
\newcommand{\INSTEJ}{\affiliation{Lancaster University, Physics Department, Lancaster, United Kingdom}}
\newcommand{\INSTFC}{\affiliation{University of Liverpool, Department of Physics, Liverpool, United Kingdom}}
\newcommand{\INSTFI}{\affiliation{Louisiana State University, Department of Physics and Astronomy, Baton Rouge, Louisiana, U.S.A.}}
\newcommand{\INSTHB}{\affiliation{Michigan State University, Department of Physics and Astronomy,  East Lansing, Michigan, U.S.A.}}
\newcommand{\INSTCE}{\affiliation{Miyagi University of Education, Department of Physics, Sendai, Japan}}
\newcommand{\INSTDF}{\affiliation{National Centre for Nuclear Research, Warsaw, Poland}}
\newcommand{\INSTFJ}{\affiliation{State University of New York at Stony Brook, Department of Physics and Astronomy, Stony Brook, New York, U.S.A.}}
\newcommand{\INSTGJ}{\affiliation{Okayama University, Department of Physics, Okayama, Japan}}
\newcommand{\INSTCF}{\affiliation{Osaka City University, Department of Physics, Osaka, Japan}}
\newcommand{\INSTGG}{\affiliation{Oxford University, Department of Physics, Oxford, United Kingdom}}
\newcommand{\INSTBB}{\affiliation{UPMC, Universit\'e Paris Diderot, CNRS/IN2P3, Laboratoire de Physique Nucl\'eaire et de Hautes Energies (LPNHE), Paris, France}}
\newcommand{\INSTGC}{\affiliation{University of Pittsburgh, Department of Physics and Astronomy, Pittsburgh, Pennsylvania, U.S.A.}}
\newcommand{\INSTFA}{\affiliation{Queen Mary University of London, School of Physics and Astronomy, London, United Kingdom}}
\newcommand{\INSTE}{\affiliation{University of Regina, Department of Physics, Regina, Saskatchewan, Canada}}
\newcommand{\INSTGD}{\affiliation{University of Rochester, Department of Physics and Astronomy, Rochester, New York, U.S.A.}}
\newcommand{\INSTHC}{\affiliation{Royal Holloway University of London, Department of Physics, Egham, Surrey, United Kingdom}}
\newcommand{\INSTBC}{\affiliation{RWTH Aachen University, III. Physikalisches Institut, Aachen, Germany}}
\newcommand{\INSTFB}{\affiliation{University of Sheffield, Department of Physics and Astronomy, Sheffield, United Kingdom}}
\newcommand{\INSTDI}{\affiliation{University of Silesia, Institute of Physics, Katowice, Poland}}
\newcommand{\INSTEH}{\affiliation{STFC, Rutherford Appleton Laboratory, Harwell Oxford,  and  Daresbury Laboratory, Warrington, United Kingdom}}
\newcommand{\INSTCH}{\affiliation{University of Tokyo, Department of Physics, Tokyo, Japan}}
\newcommand{\INSTBJ}{\affiliation{University of Tokyo, Institute for Cosmic Ray Research, Kamioka Observatory, Kamioka, Japan}}
\newcommand{\INSTCG}{\affiliation{University of Tokyo, Institute for Cosmic Ray Research, Research Center for Cosmic Neutrinos, Kashiwa, Japan}}
\newcommand{\INSTGI}{\affiliation{Tokyo Metropolitan University, Department of Physics, Tokyo, Japan}}
\newcommand{\INSTF}{\affiliation{University of Toronto, Department of Physics, Toronto, Ontario, Canada}}
\newcommand{\INSTB}{\affiliation{TRIUMF, Vancouver, British Columbia, Canada}}
\newcommand{\INSTG}{\affiliation{University of Victoria, Department of Physics and Astronomy, Victoria, British Columbia, Canada}}
\newcommand{\INSTDJ}{\affiliation{University of Warsaw, Faculty of Physics, Warsaw, Poland}}
\newcommand{\INSTDH}{\affiliation{Warsaw University of Technology, Institute of Radioelectronics, Warsaw, Poland}}
\newcommand{\INSTFD}{\affiliation{University of Warwick, Department of Physics, Coventry, United Kingdom}}
\newcommand{\INSTGH}{\affiliation{University of Winnipeg, Department of Physics, Winnipeg, Manitoba, Canada}}
\newcommand{\INSTEA}{\affiliation{Wroclaw University, Faculty of Physics and Astronomy, Wroclaw, Poland}}
\newcommand{\INSTHE}{\affiliation{Yokohama National University, Faculty of Engineering, Yokohama, Japan}}
\newcommand{\INSTH}{\affiliation{York University, Department of Physics and Astronomy, Toronto, Ontario, Canada}}

\INSTHD
\INSTEE
\INSTFE
\INSTD
\INSTGA
\INSTI
\INSTGB
\INSTFG
\INSTFH
\INSTBA
\INSTEG
\INSTDG
\INSTCB
\INSTED
\INSTEC
\INSTEI
\INSTGF
\INSTBE
\INSTBF
\INSTBD
\INSTEB
\INSTHA
\INSTCC
\INSTCD
\INSTEJ
\INSTFC
\INSTFI
\INSTHB
\INSTCE
\INSTDF
\INSTFJ
\INSTGJ
\INSTCF
\INSTGG
\INSTBB
\INSTGC
\INSTFA
\INSTE
\INSTGD
\INSTHC
\INSTBC
\INSTFB
\INSTDI
\INSTEH
\INSTCH
\INSTBJ
\INSTCG
\INSTGI
\INSTF
\INSTB
\INSTG
\INSTDJ
\INSTDH
\INSTFD
\INSTGH
\INSTEA
\INSTHE
\INSTH

\author{K.\,Abe}\INSTBJ
\author{J.\,Amey}\INSTEI
\author{C.\,Andreopoulos}\INSTEH\INSTFC
\author{L.\,Anthony}\INSTFC
\author{M.\,Antonova}\INSTEC
\author{S.\,Aoki}\INSTCC
\author{A.\,Ariga}\INSTEE
\author{Y.\,Ashida}\INSTCD
\author{Y.\,Azuma}\INSTCF
\author{S.\,Ban}\INSTCD
% ...

\collaboration{The T2K Collaboration}\noaffiliation

\date{\today}

\begin{abstract}
We report a measurement of the flux-integrated cross section for inclusive muon neutrino charged-current interactions on 
carbon. The double differential measurements are given as function of the muon momentum and angle. Relative to our 
previous publication on this topic, these results have an increased angular acceptance and higher statistics. The data 
sample presented here corresponds to $5.7 \times 10^{20}$ protons-on-target. The total flux-integrated cross section is 
measured to be $(6.950 \pm 0.662) \times 10^{-39} \text{cm}^{2}\text{nucleon}^{-1}$ and is consistent with our simulation.
\end{abstract}

\maketitle

\section{INTRODUCTION}
\label{sec:introduction}
T2K is an experiment located in Japan with the primary aim of studying neutrino oscillations \cite{T2Kexp}. It was designed to 
measure with high precision the $\nu_\mu \rightarrow \nu_\mu$ disappearance channel and to discover the 
$\nu_\mu \rightarrow \nu_e$ appearance channel. 

In addition to the oscillation measurements, T2K has an ongoing program to study neutrino interactions using the near 
detector complex in order to improve the understanding and modeling of these interactions. Results from this program, as 
exemplified by those presented in this paper, are interesting in their own right and can be used to constrain and reduce the 
systematic errors arising from cross section uncertainties in the extraction of neutrino oscillation parameters. Inclusive 
measurements provide a clear signals which are very valuable to test different models.

Previously, T2K reported the measurement of the flux-integrated double differential cross section for muon neutrino charged-
current interactions on carbon \cite{T2K-CCincl}. Since that time, many improvements have been made in the analysis. 
The results presented in this paper were obtained with more data, reduced neutrino flux uncertainties (thanks to new 
NA61/SHINE measurements \cite{NA61_09}), increased angular acceptance, reduced background contamination and a 
different unfolding method. All the improvements are described in more detail below.

The paper is organized as follows: we first summarize the experimental setup in Sec.~\ref{sec:experiment}, which contains 
the description of the off-axis beam, the near detector and the neutrino event generators used in the present analysis. The 
selection of the muon neutrino interaction samples is presented in Sec.~\ref{sec:samples} together with the summary of the 
detector systematic uncertainties. The analysis method is explained in Sec.~\ref{sec:analysis} and the results are given in 
Sec.~\ref{sec:results}.

\section{EXPERIMENTAL APPARATUS}
\label{sec:experiment}
\subsection{T2K beamline and flux prediction}

The neutrino beam used by T2K is produced at the J-PARC Laboratory in Tokai, Japan. In this process, 30 GeV/c protons 
are extracted from the main ring accelerator at J-PARC onto a graphite target, producing secondary particles consisting 
primarily of pions and kaons. The hadrons exiting the target are focused by three magnetic horns and allowed to decay in a 
decay volume. The decaying hadrons produce neutrinos (primarily of muon flavor) that continue to the near and far detectors 
while the other particles range out. Depending on the polarity of the electric current in the horns, a beam composed of 
mostly neutrinos ($\nu$-mode) or antineutrinos ($\bar{\nu}$-mode) and with energy peaked at 0.6 GeV is produced. The T2K 
beamline hardware has been described in detail elsewhere \cite{T2Kexp}.

The simulation that is used to predict the neutrino flux and its associated uncertainty is described in detail in \cite{Flux}. The 
uncertainties are dominated by the hadron production model and, to second order, by the beamline configuration. Currently, 
the uncertainty on the $\nu_\mu$ beam flux at the near detector varies from 10$\%$ to 15$\%$ depending on the neutrino 
energy. The error associated with the flux in the results presented here has been reduced with respect to that used in the 
previous analysis \cite{T2K-CCincl}, in part, because the model of hadron production from the target is tuned using the full 
2009 thin-target dataset by the NA61/SHINE experiment \cite{NA61_09}. The previous analysis used the 2007 dataset 
\cite{NA61_07}.

\subsection{The off-axis near detector}

The off-axis near detector  (ND280) is made-up of two main components, the $\pi^0$ detector (P0D \cite{P0D}) and the 
Tracker region. Both parts are contained in a metal basket box surrounded by electromagnetic calorimeters (ECal 
\cite{ECAL}) and a warm dipole magnet.  The magnet provides a 0.2 T field allowing for momentum measurement and charge 
separation.  Outside the ECal and magnet coil is the magnet flux return yoke and the side muon range detector (SMRD 
\cite{SMRD}). 

The Tracker region contains two fine-grained detectors (FGDs \cite{FGD}) sandwiched between three gas time projection 
chambers (TPCs \cite{TPC}). The TPCs contain a drift gas mixture which is ionized when a charged particle crosses it. The 
TPCs provide excellent track and momentum reconstruction. The observed energy loss in the TPCs, combined with the 
measurement of the momentum, is used for particle identification.

\begin{figure}
	\includegraphics[width=.45\textwidth]{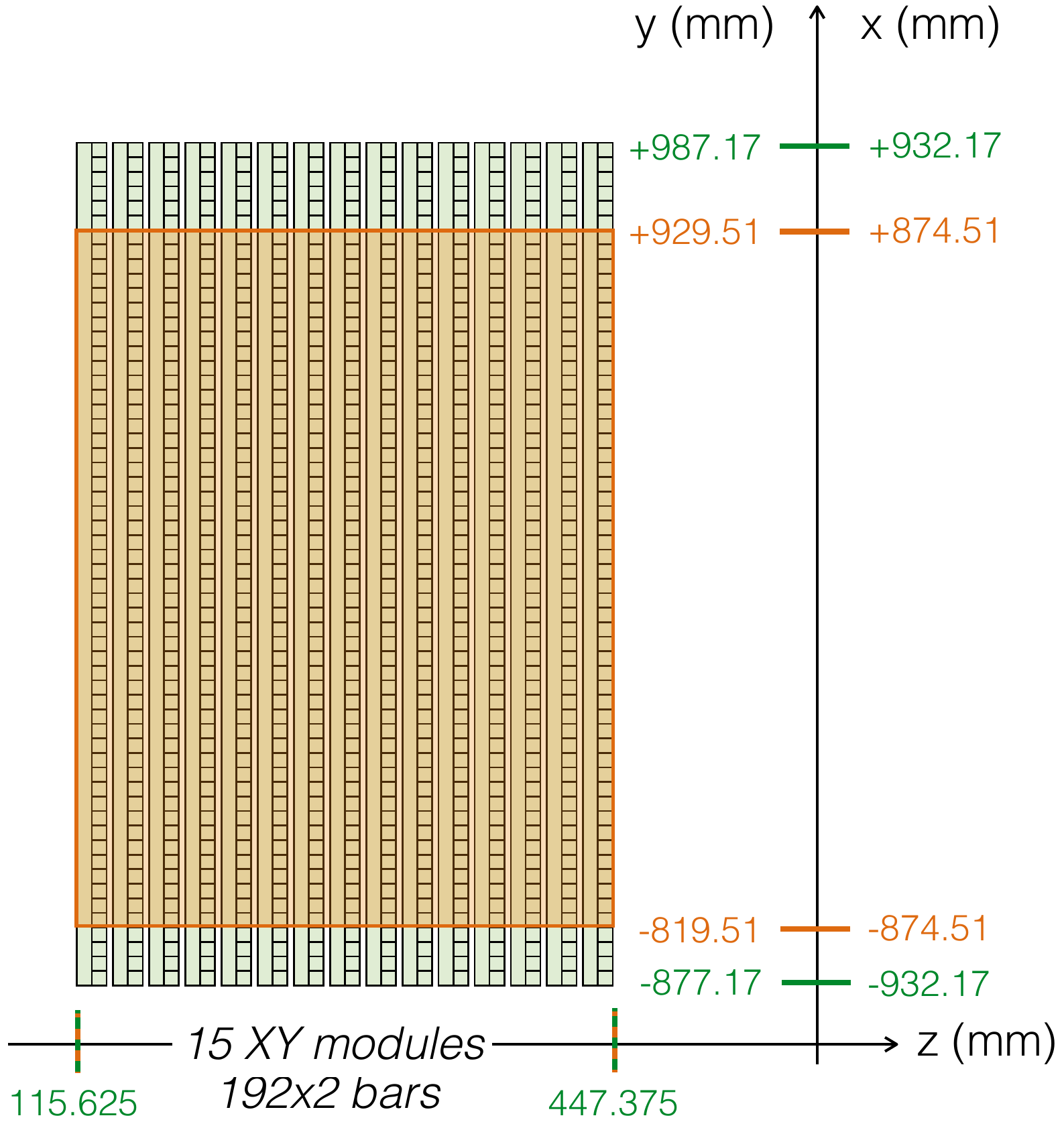}
	\caption{Schematic view of FGD1.  The orange region indicates the fiducial volume.}
	\label{fig:fiducial}
\end{figure}

The most upstream FGD (FGD1) consists of polystyrene scintillators bars, which are oriented vertically and horizontally and 
perpendicular to the beam direction. FGD1 is comprised of carbon ($86.1\%$), hydrogen ($7.4\%$) and oxygen ($3.7\%$), 
where the percentages represent the mass fraction of each element. The most downstream FGD (FGD2) is similar to FGD1 
except that the scintillators layers are interleaved with water layers. FGD1 is the active target in this analysis. The fiducial 
volume (FV) begins 58  mm inward from the lateral edges as shown in Fig.~\ref{fig:fiducial}.

The P0D region of ND280, located upstream the Tracker region, is made of layers of plastic scintillator, water, brass and 
lead. In this analysis, it is used to veto interactions happening upstream of the active target.

The SMRD consists  of 440 scintillator modules inserted in the air gaps between sections of the magnet flux return yoke. 
Horizontal (vertical) modules are composed of four (five) plastic scintillation counters.  In this analysis, the SMRD is used to 
identify and measure the range of muons at high angles with respect to the beam direction. The range provides information 
about the muon momentum.  

The ECal consist of 13 modules surrounding the inner detectors. The tracker module is covered by six modules in the sides 
(BarrelECal) and one module downstream  (DsECal). The modules are made up of plastic scintillator bars interleaved with 
lead sheets. In this analysis, the ECal is used to complement the reconstruction of the inner detectors. As with the SMRD, it 
is used to measure the range/momentum of muons escaping, from inner detectors, at high angles with respect to the beam 
direction. In addition, electromagnetic showers and minimally ionizing tracks passing through the ECal can be identified 
using a multivariate analysis quantity $R_{\text{MIP/EM}}$ determined by the features of the reconstructed clusters in 
the ECal \cite{NuE}.

\begin{figure}
	\includegraphics[width=.45\textwidth]{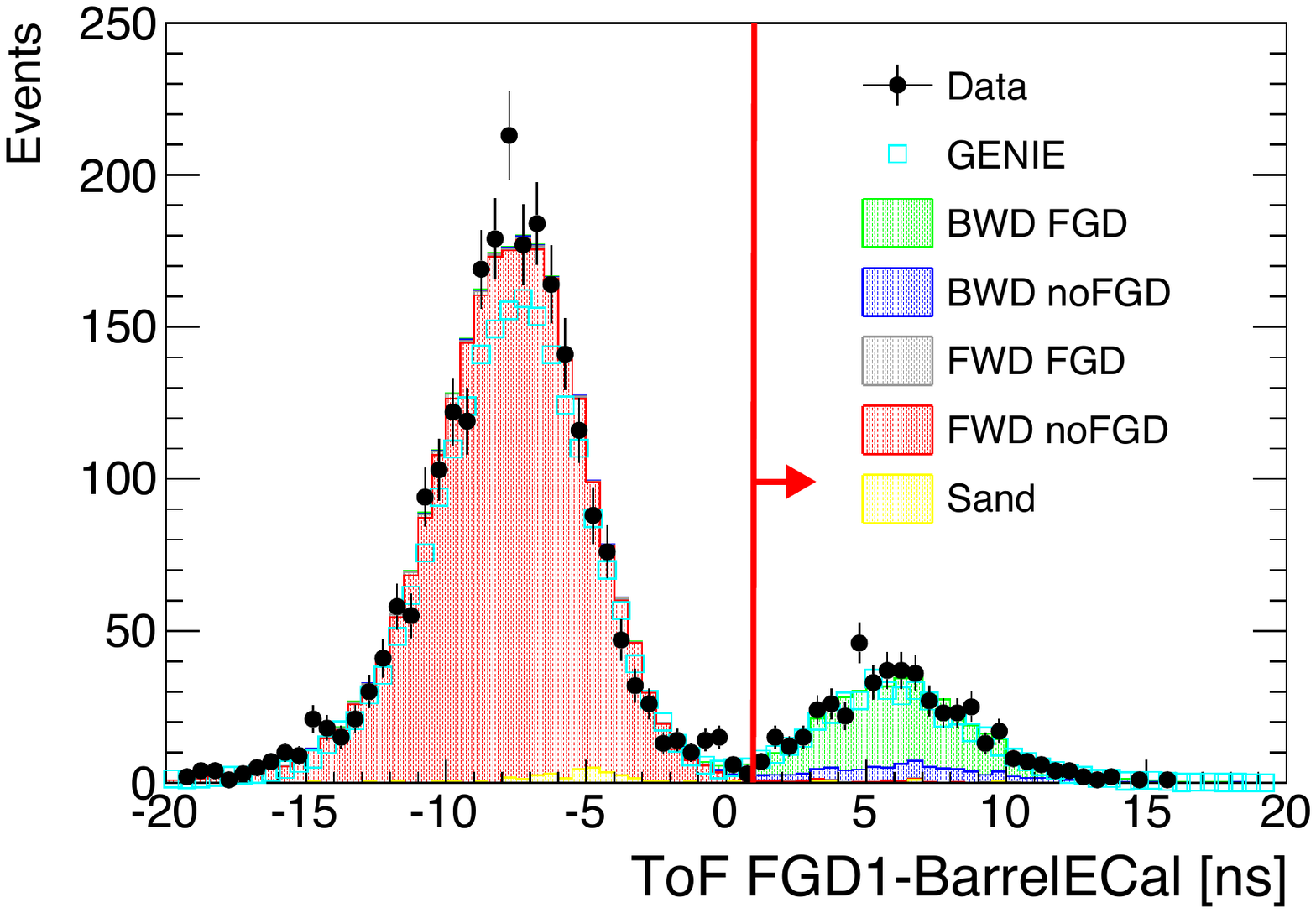}
	\caption{ToF between FGD1-BarrelECal for tracks crossing BarrelECal-TPC1-FGD1. Stacked histograms indicate the 
	prediction from NEUT of the true direction and whether the true start position is inside FGD1. Data distributions show 
	their 	statistical error bars. The region indicated by the red arrow shows tracks that are reconstructed as backward-going. 
	They are chosen that give the lowest wrong-sense fraction for each pair of detectors.}
	\label{fig:tof}
\end{figure}

In this analysis, the timing information for particles crossing the different detectors of ND280 is used for the first time. When 
a particle crosses a detector composed by scintillators, the time information from each individual hit is corrected for the light 
propagation time inside the fibers and for the time offset of each slave clock module \cite{T2Kexp}. Then, the corrected time 
and position of the hits are used to define an average time ($T$). Finally, the time of flight (ToF) variable 
(ToF$=T_{\text{X}}-T_{\text{Y}}$) between two detectors X and Y is constructed. This information is used to determine the 
direction of tracks crossing the following pairs of detectors: FGD1-FGD2, FGD1-P0D, and FGD1-BarrelECal
(see Fig.~\ref{fig:tof}).

\subsection{Event generators}
\label{sec:generators}

Two event generators, NEUT 5.3.2 \cite{NEUT} and GENIE 2.8.0 \cite{GENIE}, are used to simulate the interaction of 
neutrinos in the near detector and the effect of the nuclear medium on the produced particles. The modeling of the main 
interaction channels and their associated uncertainties is described below.

\subsubsection{Charged-current interactions without pion production}

Charged-current (CC) interactions without pion production are referred to here as charged-current quasi-elastic-like, or 
CCQE-like, interactions. The sample of such interactions is composed mainly of CCQE reactions.  However, nuclear effects 
can cause other processes to be included in this category. 

For the CCQE channel, the primary neutrino-nucleon interaction is modeled in a similar fashion by both generators. Each 
uses an implementation of the Llewellyn-Smith formalism \cite{Llewellyn} through Lorentz-invariant form factors (FFs). Both 
generators relate the vector FF to the electromagnetic FFs, for which the parametrization BBA2005 is used \cite{Bradford}. 
For the axial FF, a dipole shape with $g_A$=1.267 is used in both generators. However, the default axial mass parameter, 
$M_A$, used in each generator differs. In NEUT, $M_A = 1.21$ GeV/c$^2$, while in GENIE, $M_A = 0.99$ GeV/c$^2$. 
Finally, they use the same pseudo-scalar FF suggested by the partially conserved axial current (PCAC) hypothesis.

The majority of the CCQE interactions take place on bound nucleons. The nuclear model differs between the two generators. 
In the case of GENIE, the Bodek-Richie version of the Relativistic Fermi Gas (RFG) model is used, which incorporates short 
range nucleon-nucleon correlations \cite{Bodek0}. For NEUT, a different nuclear model is used based on the spectral 
functions from \cite{Benhar}. Moreover, NEUT includes the multi-nucleon interaction (2p2h) model from Nieves et al. 
\cite{Nieves}, as it is thought that interactions on more than one bound nucleon contribute significant strength to the signal 
relative to the single particle CCQE interaction. Pauli blocking is implemented equally in both generators (reject events with 
the momentum of the outgoing nucleon below the Fermi momentum of the nucleus).

The CCQE and 2p2h interactions are parametrized in NEUT with several target-dependent parameters (superscripts ``C" and 
``O" represent parameters for carbon and oxygen targets, respectively): the quasielastic axial mass ($M_A=1.21\pm0.3$ GeV/c$^2$), 
the binding energy ($E_b^{\text{C}}=25\pm25$ MeV and $E_b^{\text{O}}=27\pm27$ MeV), the Fermi momentum 
($p_F^{\text{C}}=217\pm30$ MeV/c and $p_F^{\text{O}}=225\pm30$ MeV/c) and the 2p2h cross-section normalization 
($\text{MEC}^{\text{C}}=1\pm1$ and $\text{MEC}^{\text{O}}=1\pm1$). The nominal values for these parameters and the 
associated uncertainties were chosen based on a study of the MINERvA and MiniBooNE datasets \cite{Callum}. Large 
uncertainties without correlations were assigned in order to cover the tensions between the two datasets and different nuclear 
models.

\subsubsection{CC interactions with pion production}

Pion production is treated differently in the two event generators. NEUT generates interactions with single pion production 
using a resonant model when $W<2$ GeV/c$^2$. Single pion production above that value and the rest of pion production 
channels are generated with a DIS model. In contrast, GENIE does not restrict the resonant model to the single pion decay 
channel. This model is switched off when $W>1.7$ GeV/c$^2$ (to avoid double counting with its DIS model). Below that 
value, the normalization of the single pion and two pions production channels from its DIS model are tuned.

Resonant pion production is based on the Rein-Sehgal model for both generators \cite{Rein}. In NEUT, the model uses 18 
resonances taking into account their interferences. The default parameters for the FFs are taken from \cite{Graczyk}. In 
contrast, GENIE incorporates 16 resonances without including interference terms and the default FFs are taken from 
\cite{Kuzmin}.

The resonant model has three parameters in NEUT: the resonant axial mass ($M_A^{RES}=0.95\pm0.15$ GeV/c$^2$), the 
normalization of the axial form factor for resonant pion production ($C_5^A=1.01\pm0.12$) and the normalization of the 
isospin non-resonant component predicted in the Rein-Sehgal model ($I_{1/2}=1.3\pm0.2$). Their nominal values and 
associated uncertainties, with no correlation assumed, were obtained by comparison with available low energy 
neutrino-deuterium single pion production data \cite{Wilkinson}.

Both NEUT and GENIE model deep inelastic scattering  using the same GRV98 PDF parametrization \cite{Gluck} including 
a Bodek-Yang correction to describe scattering at low $Q^2$. The Bodek-Yang correction differs slightly between the two 
generators, as NEUT uses \cite{Bodek1} and GENIE uses \cite{Bodek2}. An energy dependent normalisation uncertainty 
($10\%$ at 4 GeV) is used based on MINOS CC-inclusive data \cite{MINOS}.

For coherent reactions, both generators use the Rein-Sehgal model \cite{Rein2} including a correction that takes into 
account the lepton mass \cite{Rein3}. However, the implementation of the model differs slightly.  NEUT follows the 
prescriptions and data fit of pion scattering from \cite{Rein2}, leading to different cross sections for low momentum pions. 
The MINERvA experiment has reported results which are consistent with coherent pion production at $\nu$ energies around 
1 GeV \cite{MINERVA}. Considering that result, a $30\%$ normalization uncertainty in CC coherent interactions is included.

\subsubsection{Neutral-current interactions}

Neutral-current (NC) interactions affect the background prediction in this analysis. Therefore, an NC normalization 
parameter was included that scales elastic, resonant kaon and eta production, and DIS events. A 30$\%$ uncertainty is 
assigned for those channels, motivated by poor constraints from external data.

\subsubsection{Hadronization and final state interactions}

Hadron production and transport inside the nuclear medium are also simulated by the event generators. In this analysis, the 
prediction of this processes is particularly important for pions, as they contribute the main background.

The hadronization model (or fragmentation model) determines the kinematics of the primary outgoing hadrons, prior to final 
state interactions (FSI), given a particular interaction. In the high invariant mass region ($W_{\text{NEUT}}>2$ GeV/c$^2$ 
and $W_{\text{GENIE}}>3$ GeV/c$^2$), the hadronization is simulated using the PYTHIA5 and PYTHIA6 predictions 
\cite{PYTHIA} in NEUT and GENIE, respectively. These predictions are unsatisfactory near the pion production threshold. 
So, both generators include a different phenomenological description based on Koba-Nielsen-Olesen (KNO) scaling 
\cite{KNO} in the low invariant mass region. Moreover, the transition between the two regions is handled differently between 
the two generators. Specifically, GENIE includes the AGKY model \cite{AGKY} for $W<3$ GeV/c$^2$ and the transition region 
(2.3 GeV/c$^2$ $<W<$ 3 GeV/c$^2$) in which the PYTHIA model is turned on gradually.

In GENIE, several parameters affect pion kinematics. In particular, for single pion states four parameters are notable: the 
nucleon $x_F$ ($p_T^2$), PDFs for $N\pi$ hadronic states, the nuclear formation zone, and the pion angular distribution in 
$\Delta$ resonant pion production. Their nominal values and associated uncertainties are estimated based on 
recommendations from the GENIE Collaboration \cite{GENIE}. These parameters are treated as uncorrelated.

Near an energy of 1 GeV, pions immersed in a highly dense nuclear medium are very likely to interact.  Both generators 
simulate pion FSI using the intra-nuclear cascade approach, though they use different predictions for the interaction 
probabilities.  In the case of NEUT, pion interaction probabilities are dependent on the momentum of the pion: if 
$p_\pi<500$ MeV/c, NEUT uses a density dependent model \cite{Salcedo} and if $p_\pi>500$ MeV/c the probabilities are 
extracted from pion-nuclear scattering experiments \cite{Ashery}. GENIE uses a model called INTRANUKE hA which 
extracts the interaction probabilities from several experiments up to 300 MeV/c, while for higher energies it is based on the 
CEM03 predictions \cite{Mashnik}. The uncertainties associated with the pion interaction probabilities and their correlations 
are estimated using the same methodology as in \cite{LongOA}.

\section{\texorpdfstring{$\nu_\mu$}{nu} CC SAMPLES}
\label{sec:samples}
This analysis uses data collected in $\nu$-mode between November 2010 and May 2013. The total sample comes from 
$5.7\times10^{20}$ protons on target (POT), which is a factor of five larger than that used in the similar previously published 
analysis from T2K \cite{T2K-CCincl}.

Simulated Monte Carlo (MC) interactions within the ND280 subdetectors  and magnet were generated using both NEUT and 
GENIE.  The background interactions in the materials surrounding ND280, so-called sand interactions, were generated 
using NEUT. Both interactions in ND280 and in the surrounding material were generated using the same neutrino beam 
simulation, detector simulation and reconstruction.

In this analysis, events containing muons emanating from interactions that occur in the fiducial volume (FV) of FGD1 are 
selected. These events are candidate $\nu_\mu$ CC interactions. The events within this sample that are true $\nu_\mu$ CC 
events belong to the category referred to here as $\nu_\mu$CC-$\mu$.  

Background events in the initial selection include: interactions not happening in the FV (either inside or outside the magnet 
volume, referred to as `out FV' and `sand~$\mu$', respectively); interactions happening in the FV but not actually a $
\nu_\mu$ CC event, referred to as no$\nu_\mu$CC; or being $\nu_\mu$ CC but where the muon candidate track is not the 
outgoing muon, herein called $\nu_\mu$CC-no$\mu$. 

The cross-section results presented here are based on the kinematics of the outgoing muon. Specifically, the results are 
given as a function of the muon momentum, $p_\mu$, and the cosine of the muon emission angle with respect to the 
neutrino direction, $\cos \theta_\mu$. The event selection criteria and performance, as well as the systematic uncertainties 
associated with the detector response are described below.

\subsection{Event selection}

In previous T2K work on this topic, the analysis was optimized to select forward-going muons originating from FGD1 and 
making a long track (at least 19 clusters as described in section~\ref{sec:fwd}) through TPC2, which is downstream of 
FGD1 \cite{T2K-CCincl}. The current work aims to include the so-called €œhigh-angle€ tracks which miss or barely cross the 
TPCs, as well as long backward-going tracks in TPC1 (upstream of FGD1). The addition of backward-going muon 
candidates in the event selection is possible only with the introduction of timing information correlated between 
subdetectors.

\begin{figure}
	\includegraphics[width=.45\textwidth]{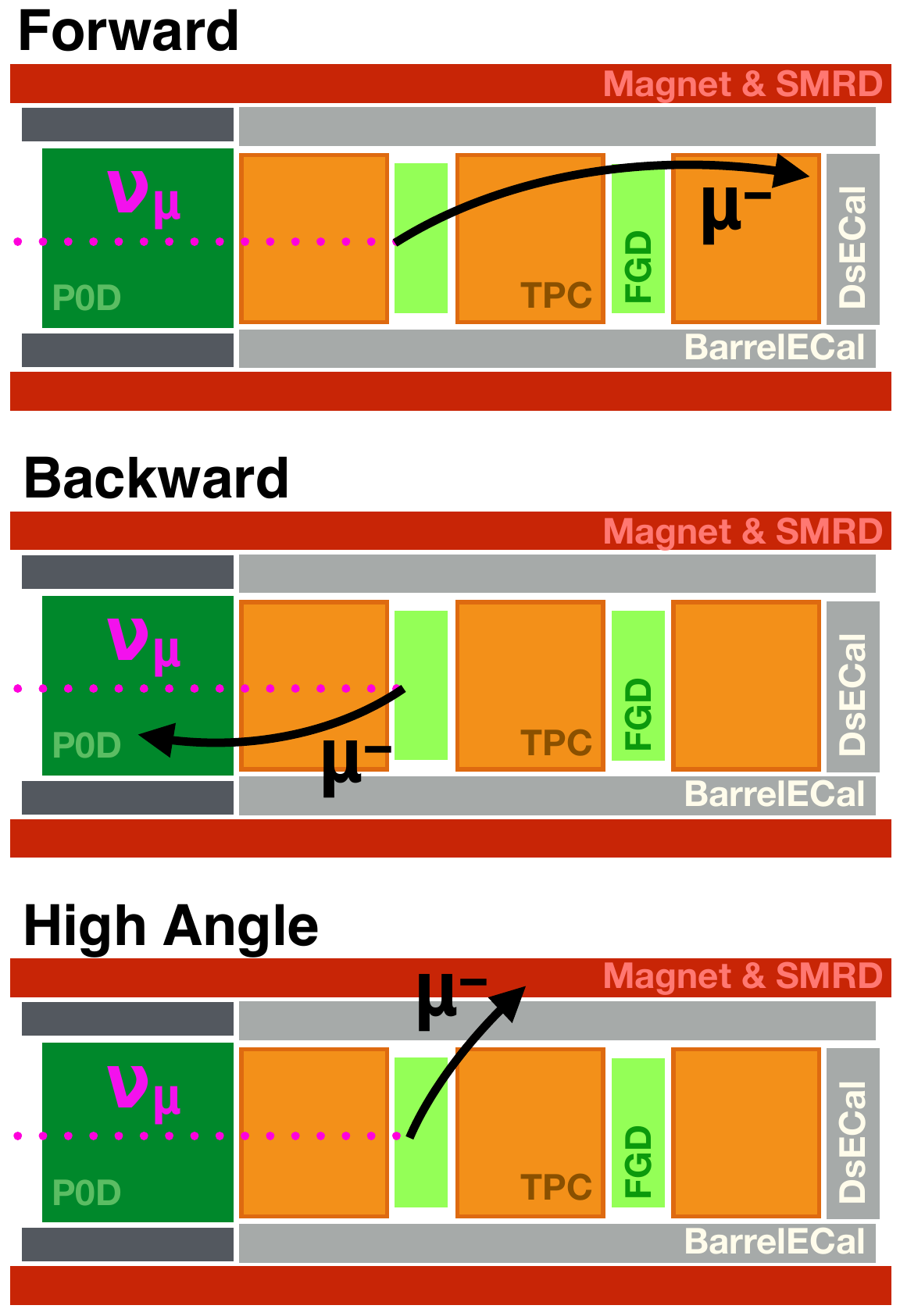}
	\caption{Schematic representation of the regions of interest for each selection.}
	\label{fig:sketch}
\end{figure}

In this analysis, events are broken into samples according to the muon direction.  If the muon candidate in the event goes 
forward (in the direction downstream of FGD1 into TPC2), the event is part of the forward (FWD) sample. If the muon goes 
backward (in a direction upstream of FGD1 into TPC1), the event is part of the backward (BWD) sample. Similarly, if the 
muon candidate in the event is at a high angle in the forward or backward direction, the event is categorized as high-angle 
forward (HAFWD) or high-angle backward (HABWD), respectively.  In the FWD/BWD selections, the muon candidate must 
have long TPCs segments, while tracks with short or no TPC segment are used in the HAFWD/HABWD (see 
Fig.~\ref{fig:sketch}).

For events to be considered in this analysis, they must occur within the time window of one of the 8 beam bunches per 5 $
\mu s$ spill RF structure of the beam. The full spill is required to be of good quality. Events are resolved in time by bunch 
and then processed. Given the beam intensity for these runs, the frequency of multiple neutrino interactions happening in 
the same beam spill (so-called pile-up events) is very low.  This is ignored in the sample selection and included in the 
systematic error treatment.

In order to avoid having multiple muon candidates, the analysis looks for candidates sequentially in the different event 
selections. The ordering for this process is FWD, BWD, and then the high angle categories. FWD and BWD have a higher 
priority than the high angle categories because the muon PID from the TPCs is more accurate than in the ECals. The 
FWD(HAFWD) selection has a higher priority than the BWD(HABWD) because forward-going muon happen much more 
often than backward-going ones.

Additionally, two control regions are selected to constrain neutral current event rates and pion final state interactions. The 
control regions are non-signal regions of phase space close enough to the signal region that the backgrounds are similar to 
that in the signal region.  The backgrounds used in the model are tuned using the data observed in the control regions.  The 
control region selection is described in section~\ref{sec:controlregions}.

\subsubsection{Forward selection}
\label{sec:fwd}

The selection criteria for the FWD sample are very similar to those used previously, though some further optimization has 
been performed. The cuts used to extract the FWD sample are described below.

\begin{itemize}
	\item \textbf{Quality and FV}: This selection considers negatively charged tracks originating in the FGD1 FV which 
	have TPC track segments containing more than 18 clustered hits in the TPC. If multiple tracks satisfy these criteria, the 
	muon candidate is the one with highest momentum and going forward (by timing). In order to reduce the contamination 
	from events occurring outside the FV, tracks starting in the most upstream layer of FGD1 are rejected.
	\item \textbf{Muon PID}: This cut is applied to the muon candidate using discriminator functions calculated for muon, 
	pion and proton hypotheses based on the energy loss and momentum measurement of the TPC. These functions are 
	the same as used in the previous analysis \cite{T2K-CCincl}. This cut rejects protons, pions and low momentum 
	electrons (below 500 MeV/c). Moreover, two new PID cuts below have been developed in order to reduce the pion 
	contamination of this sample (which is the main background in this analysis).	
	\begin{itemize}
		\item \textit{Muon FGD2 PID}: High energy pions are more likely to stop in FGD2 than muons. Therefore, it is 
		required that the muon candidate leave the FGD2 active volume with a momentum above 280 MeV/c. This is 
		expected to reduce the pion contamination by $15\%$ while leading to a loss of $0.3\%$ of the muons.		
		\item \textit{Muon ECal PID}: For tracks entering the BarrelECal or DsECal modules, the multivariate analysis 
		quantity $R_{\text{MIP/EM}}$ (based on the features of the reconstructed clusters in the ECal \cite{NuE}) 
		is used. These tracks must have $R_{\text{MIP/EM}}<15$, which is estimated to reduce the pion contamination  
		by  $7\%$ while removing $0.3\%$ of the muons.
	\end{itemize}		
	\item \textbf{Veto}: One of the main backgrounds in this analysis are interactions happening outside the FV. This 
	contamination can be reduced further by using the two cuts described below:
	\begin{itemize}
		\item \textit{Upstream background veto}: Due to reconstruction failures and multiple scattering, a reconstructed 
		track can be broken into two unmatched segments. One of those can have its beginning in the FV, mimicking an 
		interaction that originates in the FV. In the previous analysis, such events were rejected if the second highest 
		momentum track started more than 150 mm upstream of the muon candidate. This cut was found to be too 
		restrictive because it removed events with a forward going muon and a second particle going backward. In the 
		current analysis, the ratio between the momentum of the muon candidate and the other track is used. Ideally, if 
		the muon candidate is a broken track, this ratio should be bigger than one since the first segment of the track 
		has a higher momentum than the second segment. Therefore, the distance between both tracks, or segments, as 
		well as their momentum ratio are used. Cut values are chosen that give the highest purity times efficiency.		
		\item \textit{Broken track cut}: This cut rejects events where the reconstruction procedure mistakenly breaks a 
		single track into two tracks where the first is a FGD1 segment and the second is reconstructed to begin in the last 
		layers of FGD1 and goes through the downstream TPC module. In this mis-reconstruction pathology, the second 
		track is considered a muon candidate. For such events, the start position of muon candidate track is within the 
		two most downstream layers of FGD1. The broken track cut rejects these events by requiring that there be no 
		reconstructed track with only a FGD1 segment when the start position of the muon candidate is in one of the last 
		two layers of FGD1.		
	\end{itemize}	
\end{itemize}

\begin{figure}
	\includegraphics[width=.45\textwidth]{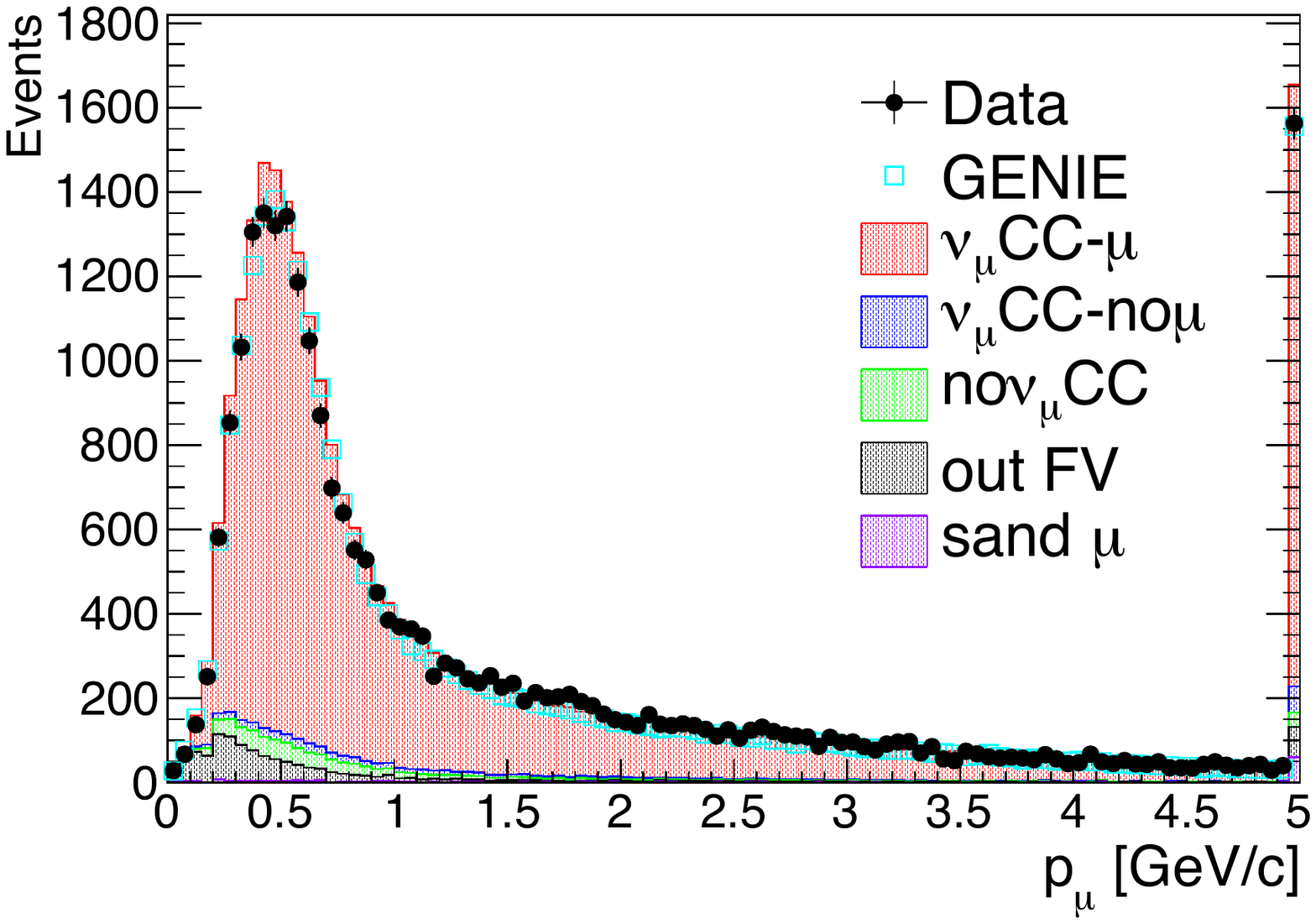}
	\includegraphics[width=.45\textwidth]{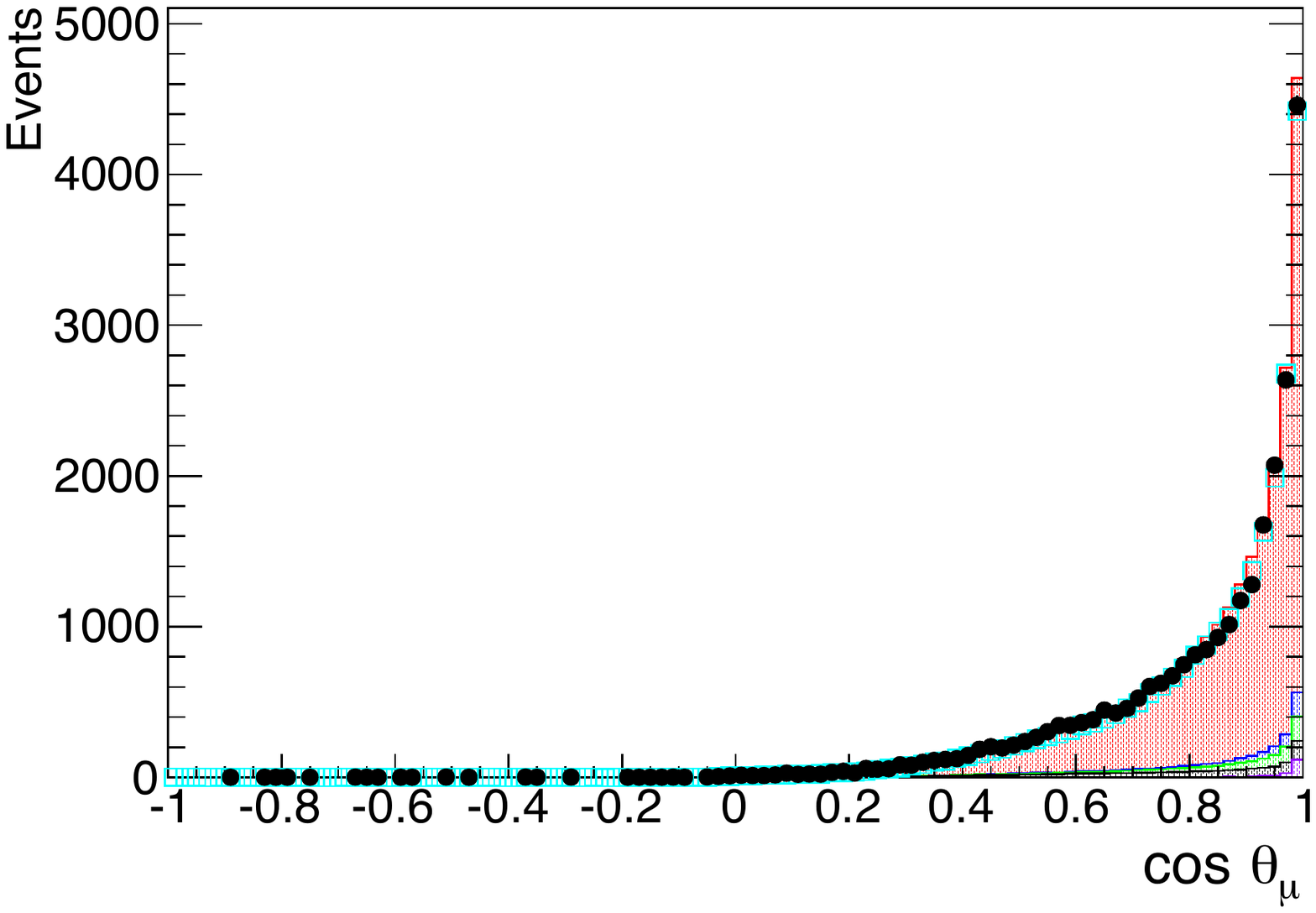}
	\caption{Momentum (top) and cosine of emission angle (bottom) for the muon candidate when all selection criteria are 
	fulfilled in the FWD selection. Stacked histograms indicate different reaction types predictions from NEUT. Empty 
	rectangles indicate the prediction from GENIE. Data distributions show their statistical error bars.}
	\label{fig:fwd_cc4pi}
\end{figure}

Fig.~\ref{fig:fwd_cc4pi} shows the reconstructed kinematics for muon candidates in the FWD sample in the data together with 
the prediction from NEUT and GENIE.

\subsubsection{Backward selection}

The selection criteria for the BWD sample are described below:

\begin{itemize}
	\item \textbf{Quality and FV}: This selection considers negatively charged tracks originating in the FGD1 FV which 
	have TPC track segments containing more than 18 clusters. If the event contains multiple tracks of this type, the muon 
	candidate is the one with highest momentum and backward sense (by timing). In order to reduce the 
	contamination from events occurring outside the FV, tracks starting in the most upstream layer of FGD1 are rejected.
	\item \textbf{Muon PID}: For muon candidates in the BWD sample, the PID is  based entirely on the energy loss in the 
	TPC. The value of the cut applied is the same as that in the FWD selection. However, in this angular region the 
	electron contamination is very low and the discriminator function used to reduce the low momentum electrons is not 
	applied.
\end{itemize}

\begin{figure}
	\includegraphics[width=.45\textwidth]{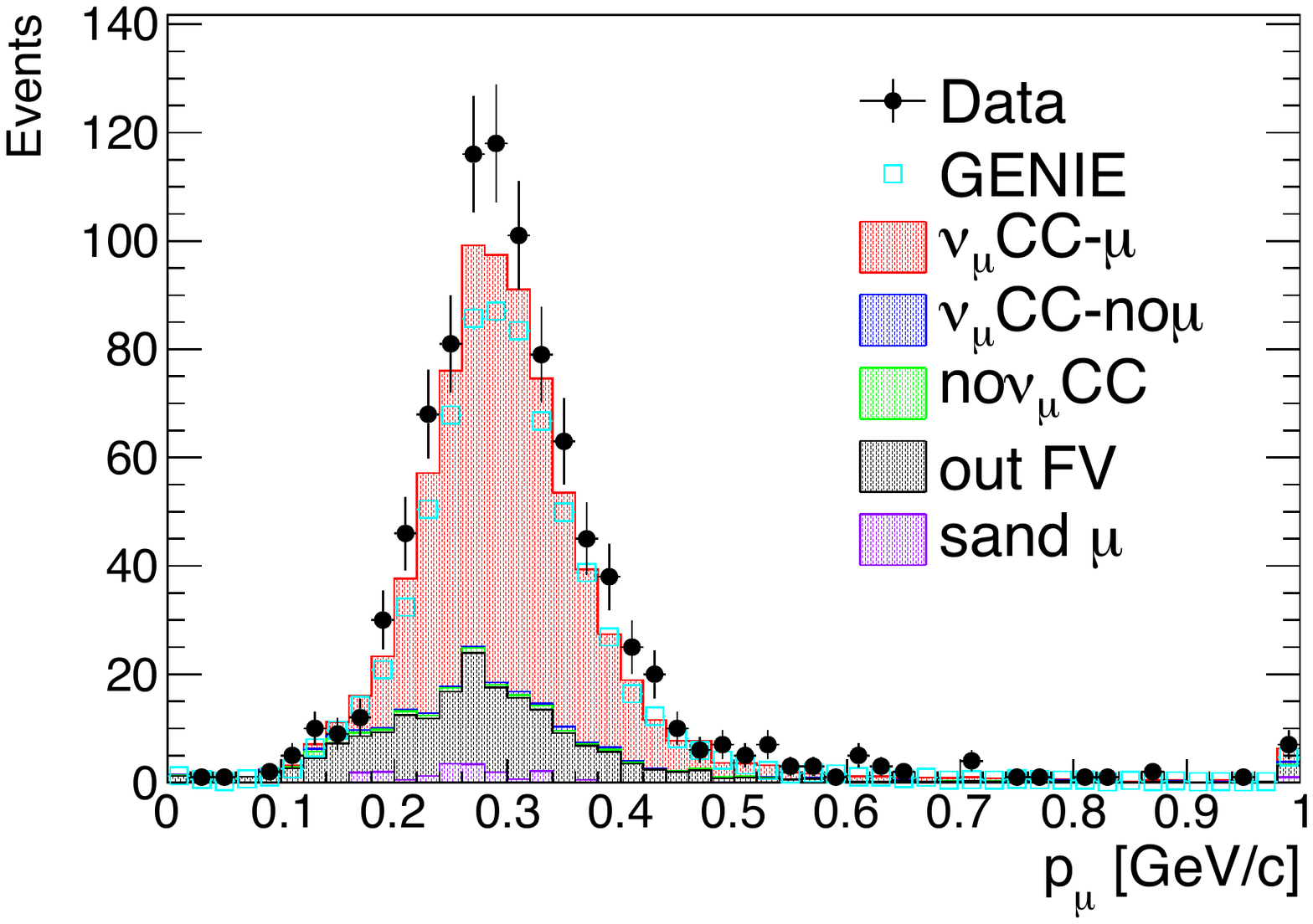}
	\includegraphics[width=.45\textwidth]{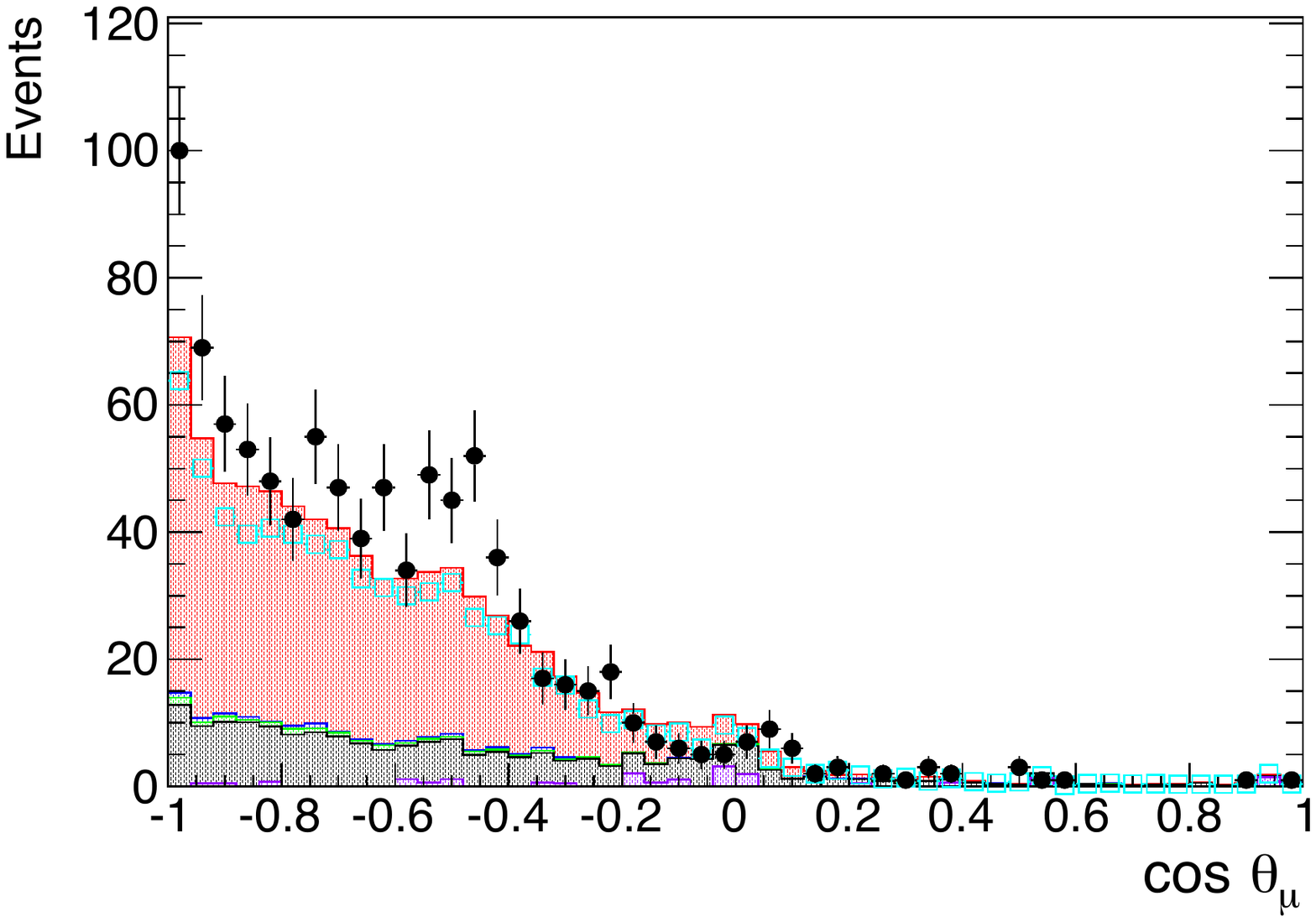}
	\caption{Momentum (top) and cosine of emission angle (bottom) for the muon candidate when all selection criteria are 
	fulfilled in the BWD selection. Stacked histograms indicate different reaction types predictions from NEUT. Empty 
	rectangles indicate the prediction from GENIE. Data distributions show their statistical error bars.}
	\label{fig:bwd_cc4pi}
\end{figure}

Fig.~\ref{fig:bwd_cc4pi} shows the reconstructed kinematics for muon candidates in the BWD sample in the data together with 
the prediction from NEUT and GENIE.

\subsubsection{High Angle selection}

In the selection for the high angle samples (HAFWD and HABWD), the muon candidates are mostly (or all) contained in the 
FGD1, ECal and SMRD subdetectors. A detailed explanation of the selection criteria is shown below.

\begin{itemize}
	\item \textbf{Quality and FV}: High angle tracks starting in FGD1 FV and stopping either in SMRD or BarrelECal are 
	considered. The stopping requirement is needed in order to compute the momentum of the track by range. The 
	contamination from events occurring outside the FV is reduced by rejecting tracks starting in the most upstream or 
	downstream layers of FGD1.
	\item \textbf{Muon PID}: The TPC PID information is not reliable for high angle tracks since they have no (or short) 
	TPC segments. The SMRD and BarrelECal information forms the basis of the high angle track PID. Tracks that reach 
	the SMRD in the HAFWD sample are good muon candidates ($\sim$1200 tracks). In the HABWD sample, most tracks 
	reaching the SMRD come from out of the FV. Consequently, tracks reaching the SMRD in the HABWD sample are 
	rejected ($\sim$70 tracks). Tracks not reaching the SMRD and stopping in the BarrelECal region of the detector 
	($\sim$4250 and $\sim$1250 tracks for HAFWD and HABWD respectively) are considered as muon candidates if the 
	multivariate analysis quantity $R_{\text{MIP/EM}}<0$. Besides, we reduce the contamination of protons rejecting 
	events that release high amount of energy in short distances within the BarrelECal.
	\item \textbf{Veto}: The upstream background veto, introduced in the FWD selection, is used for the high angle 
	samples. For this veto, the distance and momentum ratio relation was optimized for forward going and backward going 
	candidates independently.
\end{itemize}

\begin{figure}
	\includegraphics[width=.45\textwidth]{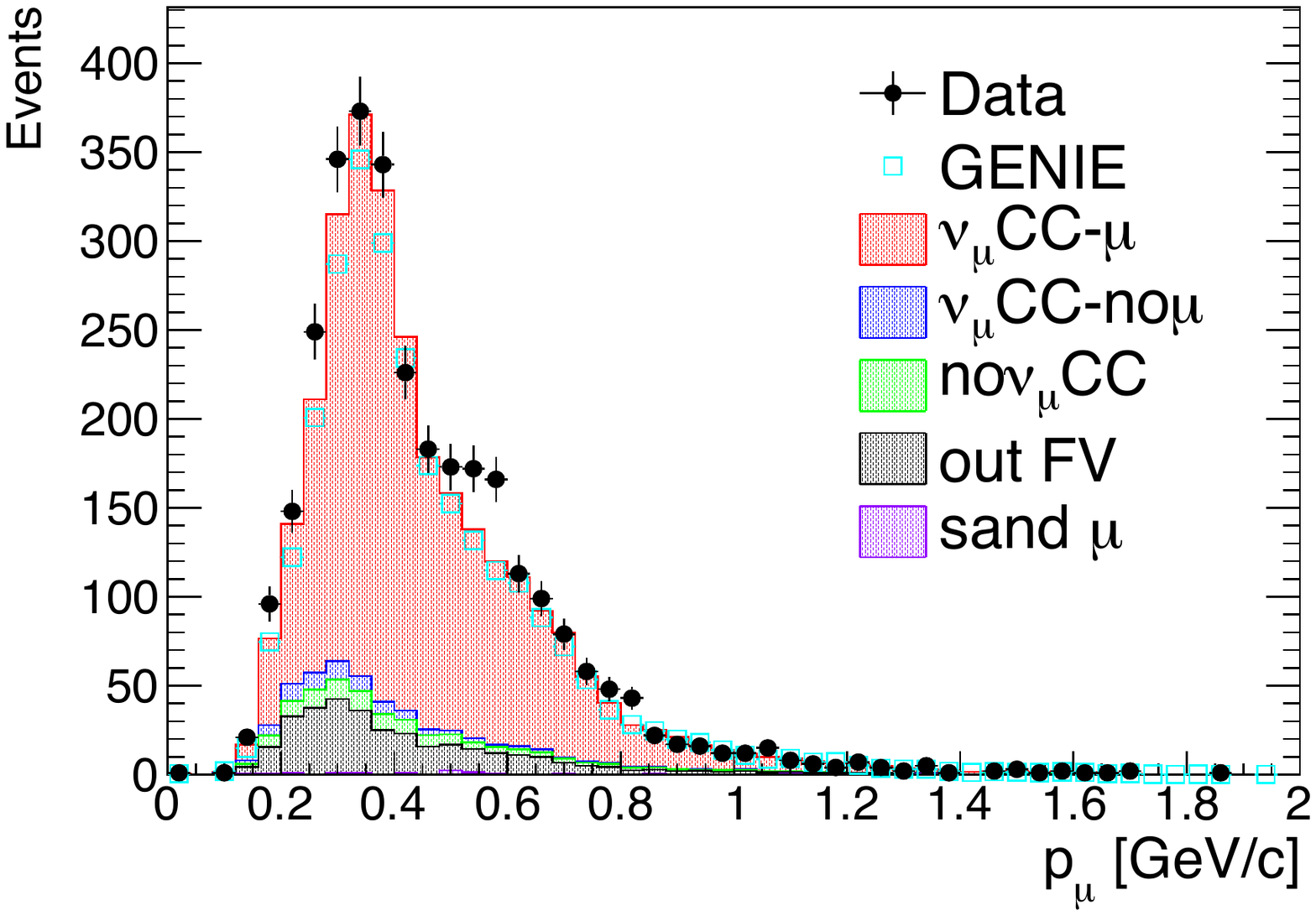}
	\includegraphics[width=.45\textwidth]{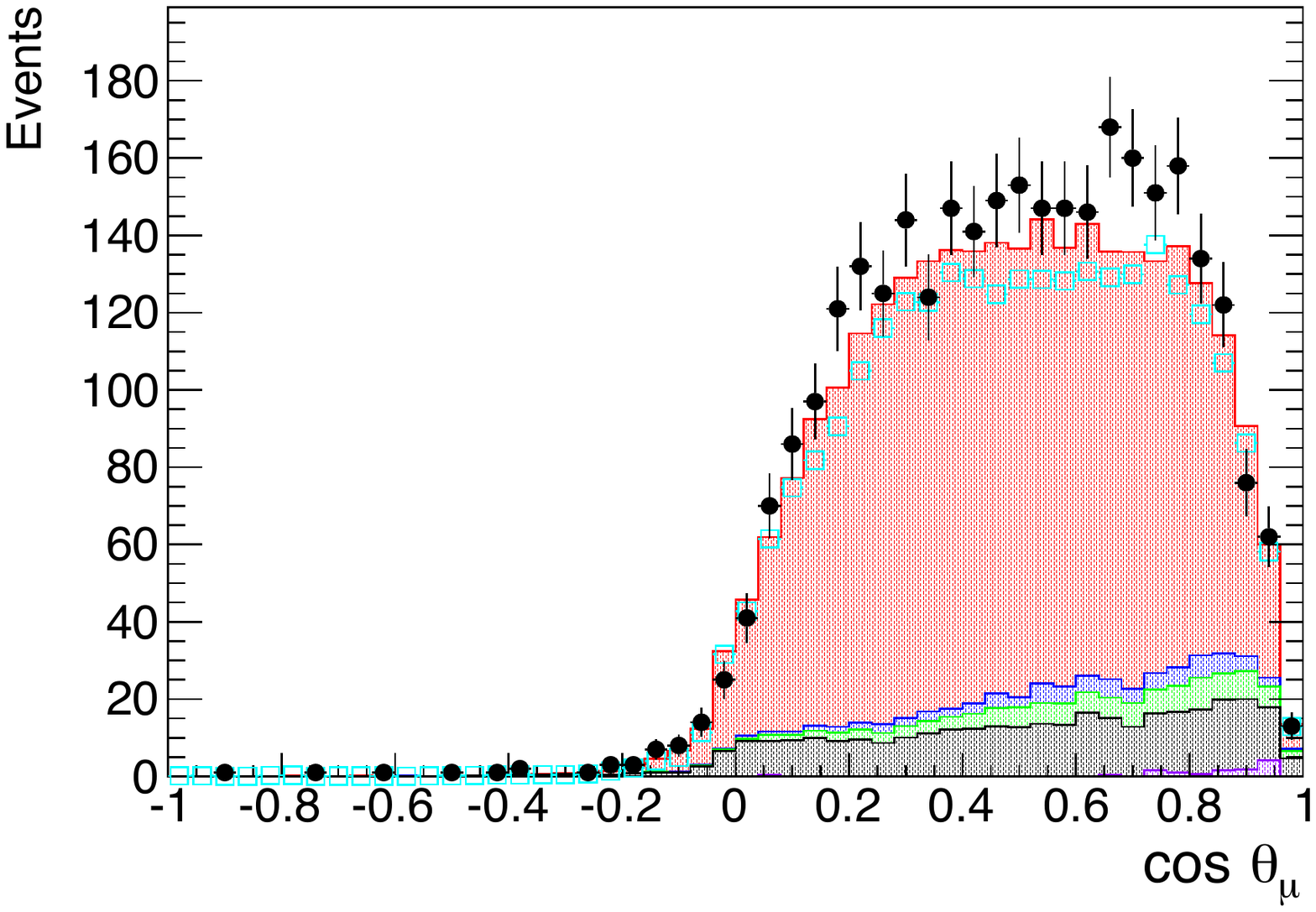}
	\caption{Momentum (top) and cosine of emission angle (bottom) for the muon candidate when all selection criteria are 
	fulfilled in the HAFWD selection. Stacked histograms indicate different reaction types predictions from NEUT. Empty 
	rectangles indicate the prediction from GENIE. Data distributions show their statistical error bars.}
	\label{fig:hafwd_cc4pi}
\end{figure}

\begin{figure}
	\includegraphics[width=.45\textwidth]{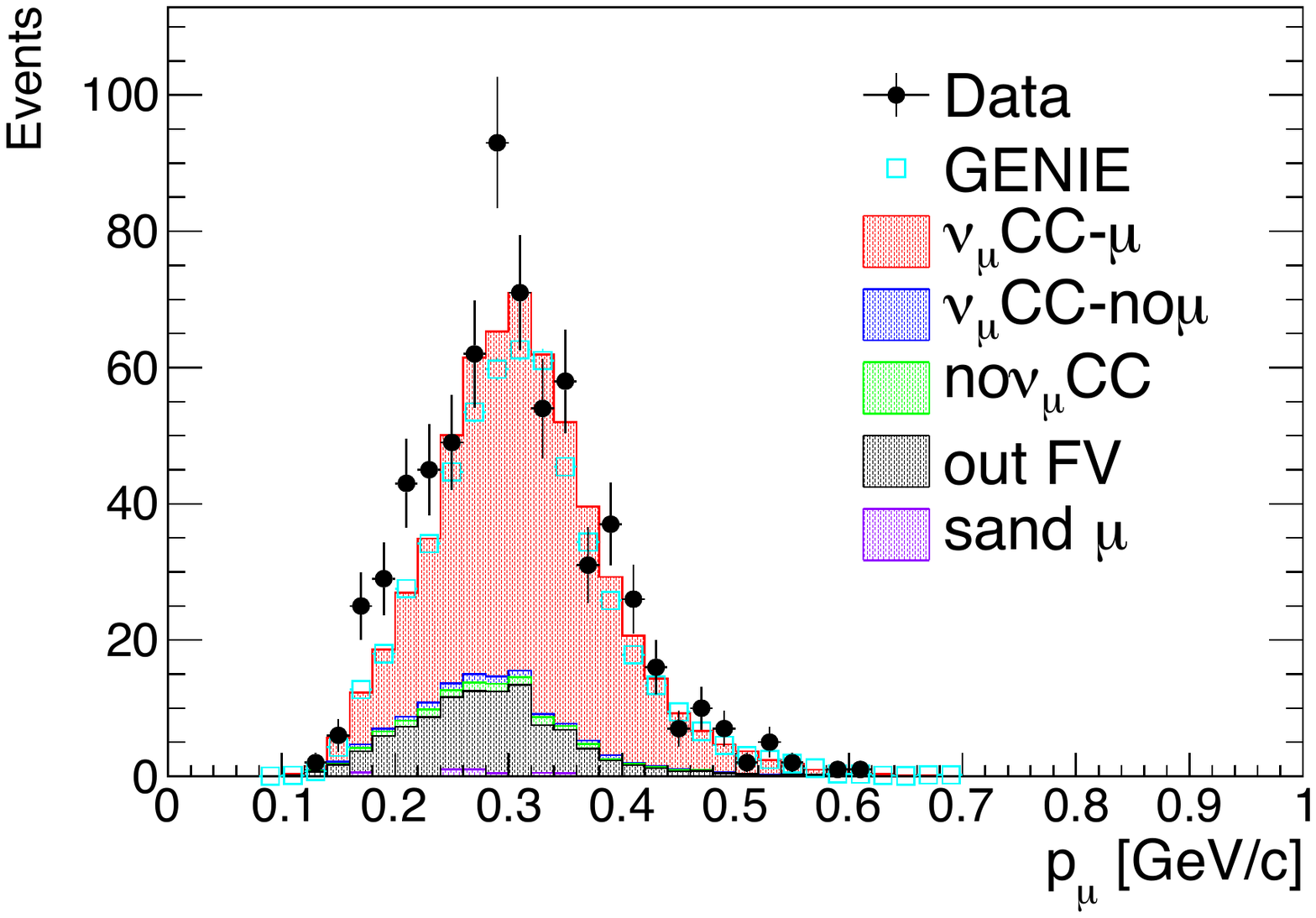}
	\includegraphics[width=.45\textwidth]{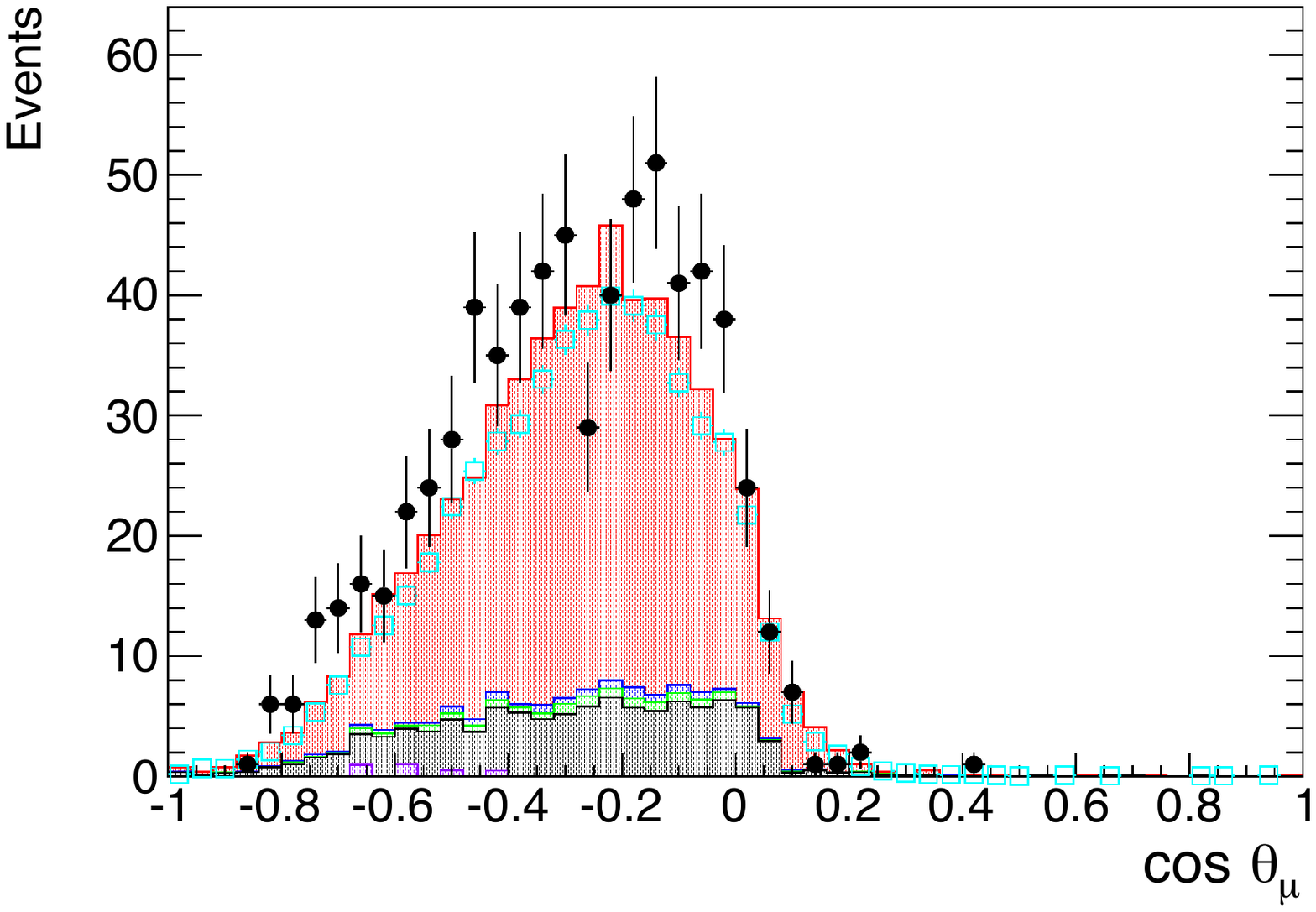}
	\caption{Momentum (top) and cosine of emission angle (bottom) for the muon candidate when all selection criteria are 
	fulfilled in the HABWD selection. Stacked histograms indicate different reaction types predictions from NEUT. Empty 
	rectangles indicate the prediction from GENIE. Data distributions show their statistical error bars.}
	\label{fig:habwd_cc4pi}
\end{figure}

Fig.~\ref{fig:hafwd_cc4pi} and Fig.~\ref{fig:habwd_cc4pi} show the reconstructed kinematics for the muon candidates in the 
HAFWD and HABWD samples in the data together with the prediction from NEUT and GENIE.

\subsubsection{Control regions selection}
\label{sec:controlregions}

As mentioned earlier, uncertainties associated with the modeling of backgrounds and pion kinematics, neutral current 
normalization and pion final state interactions can be minimized using control regions.  The backgrounds used in the model 
are tuned using the data observed in the control regions.

\begin{figure}
	\includegraphics[width=.45\textwidth]{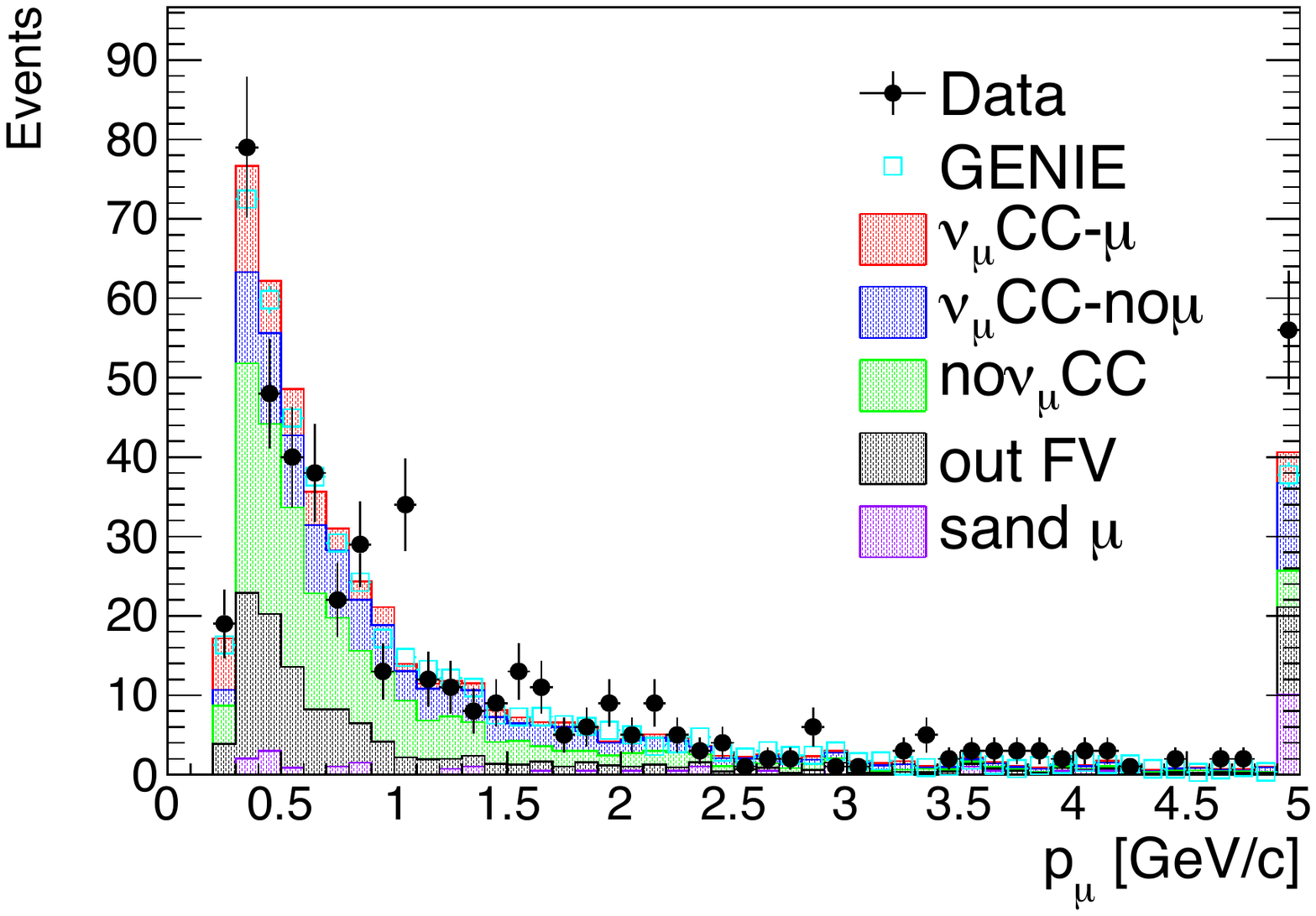}
	\includegraphics[width=.45\textwidth]{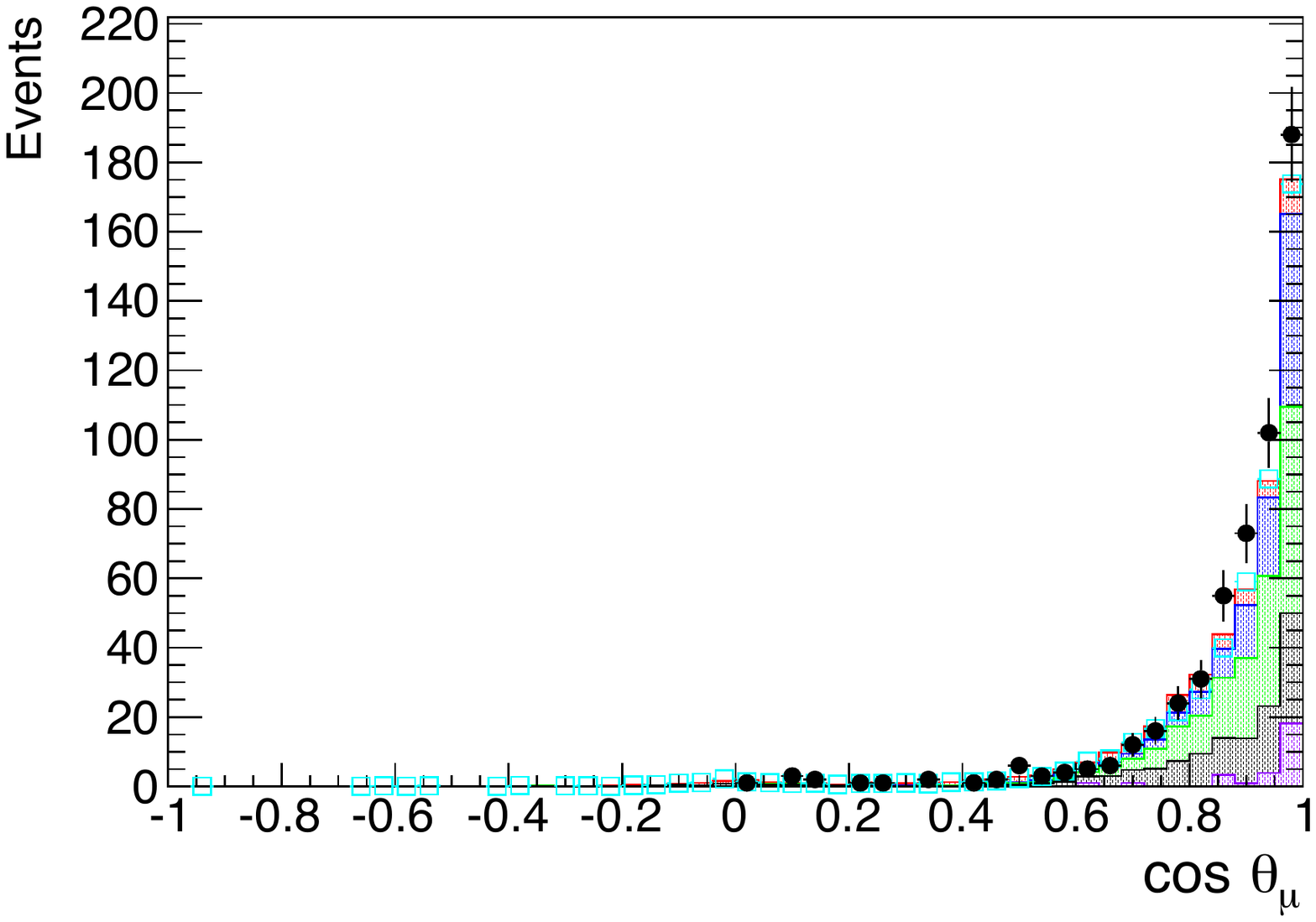}
	\caption{Momentum (top) and cosine of emission angle (bottom) for the pion candidate when all selection criteria are 
	fulfilled in the CSFGD2 selection. Stacked histograms indicate different reaction types predictions from NEUT. Empty 
	rectangles indicate the prediction from GENIE. Data distributions show their statistical error bars.}
	\label{fig:controlregion1}
\end{figure}

\begin{figure}
	\includegraphics[width=.45\textwidth]{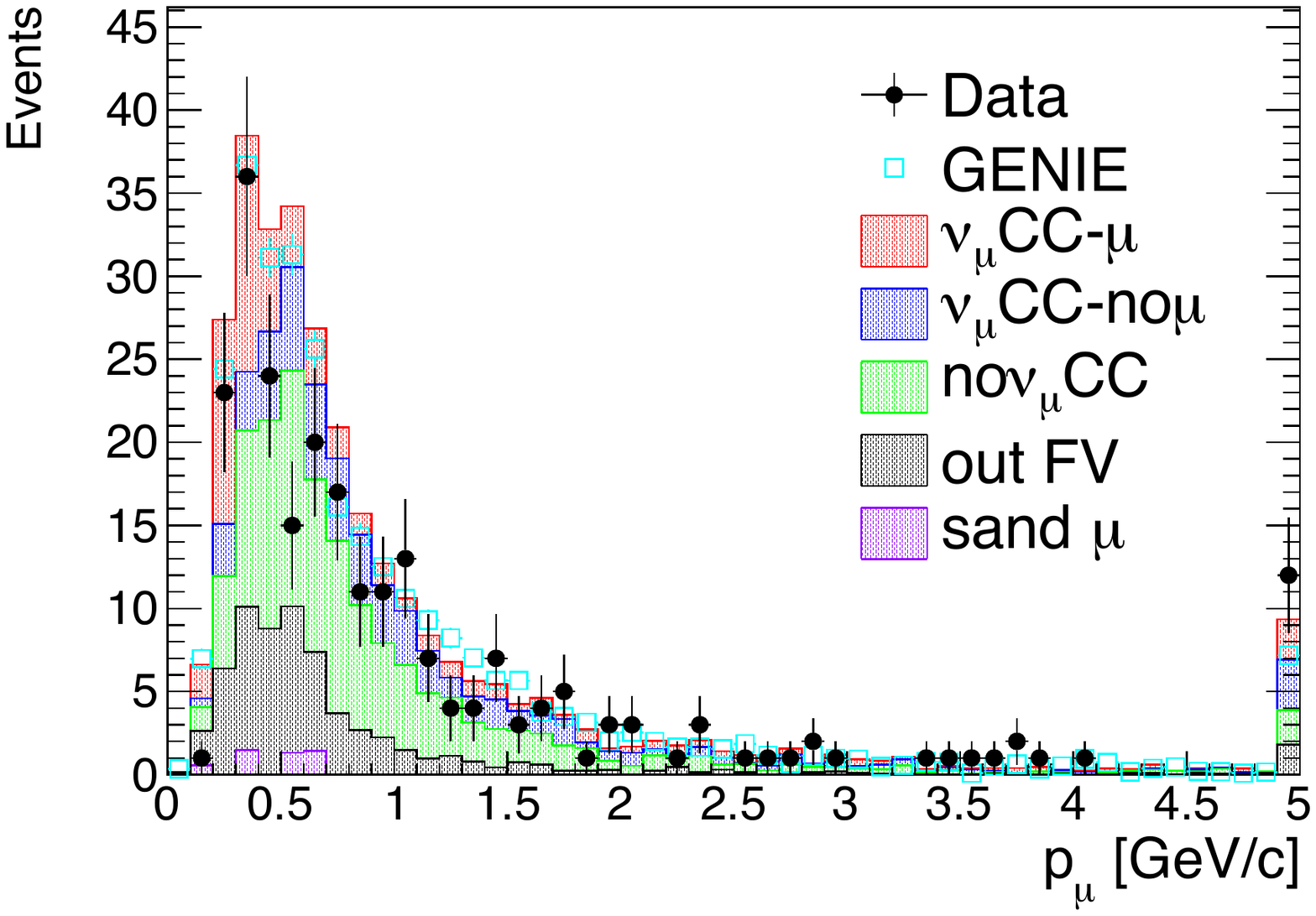}
	\includegraphics[width=.45\textwidth]{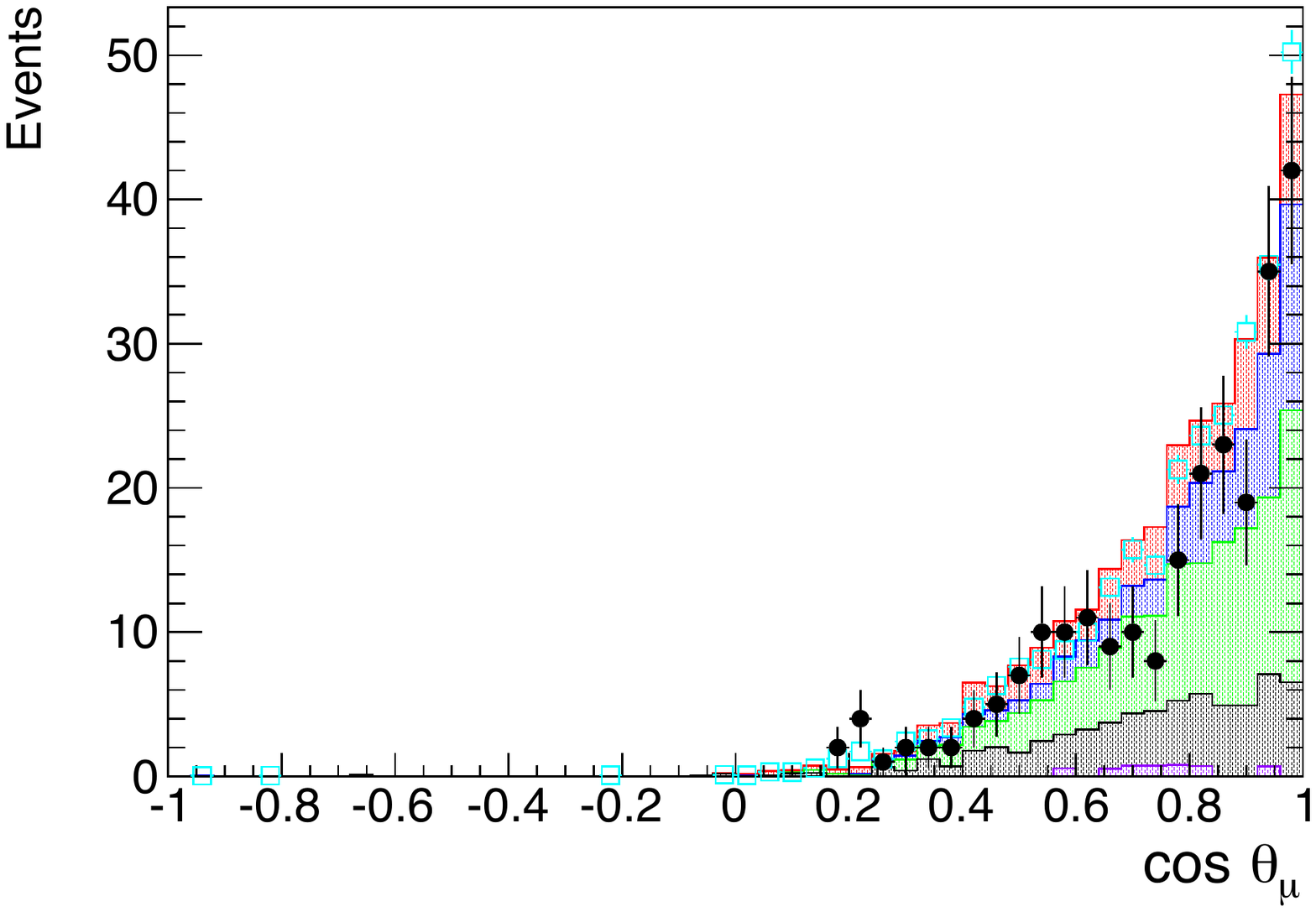}
	\caption{Momentum (top) and cosine of emission angle (bottom) for the pion candidate when all selection criteria are 
	fulfilled in the CSECAL selection. Stacked histograms indicate different reaction types predictions from NEUT. Empty 
	rectangles indicate the prediction from GENIE. Data distributions show their statistical error bars.}
	\label{fig:controlregion2}
\end{figure}

Events that do not fulfill the muon ECal PID and muon FGD2 PID in the FWD selection constitute the control region 
samples, CSECAL and CSFGD2, respectively. Fig.~\ref{fig:controlregion1} and Fig.~\ref{fig:controlregion2} show the 
reconstructed kinematics for muon candidates in the control region samples in data as well as the expectation from NEUT 
and GENIE. A relative good agreement is observed within systematic uncertainties, which are particularly large in these 
samples (mainly affected by detector response). The main contribution (70$\%$) in both control samples are negative pions 
formed in NC or CC deep inelastic interactions. The fraction of signal events in each control sample is below 20$\%$. 

\subsection{Signal and background compositions of the selection}

Table~\ref{table:reduction_per} summarizes how each step in the selection affects the number of events and purity in each 
sample in both data and MC. Both the PID and veto cuts play a significant role in increasing the purity in each sample.

\begin{table*}
	\begin{ruledtabular}
		\begin{tabular}{ c | c | c | c | c | c | c | c | c }
			\multicolumn{1}{c|} {Cut} & \multicolumn{2}{c|}{FWD} & \multicolumn{2}{c|}{BWD} & \multicolumn{2}{c|}{HAFWD} & \multicolumn{2}{c}{HABWD} \\
			\hline
			\multicolumn{1}{c|} {} & \multicolumn{1}{c|}{DATA} & \multicolumn{1}{c|}{NEUT} & \multicolumn{1}{c|}{DATA} & \multicolumn{1}{c|}{NEUT} & \multicolumn{1}{c|}{DATA} & \multicolumn{1}{c|}{NEUT} & \multicolumn{1}{c|}{DATA} & \multicolumn{1}{c}{NEUT} \\
			\hline
			Quality & 
			82155 & 81222 & 
			1861  & 1050 & 
			7225  & 7121 & 
			1582  & 1566 \\
			& & \textbf{32.3} & & \textbf{58.5} & & \textbf{41.8} & & \textbf{48.9} \\
			\hline
			FV &   
			50519 & 51648 & 
			1165  & 1025 &
			5669  & 5764 &
			1356  & 1360 \\
			& & \textbf{48.7} & & \textbf{58.8} & & \textbf{49.2} & & 54.1 \\
			\hline
			$\mu$ PID & 
			29140 & 29750 &
			940   & 799   &
			3712  & 3487 &
			779   & 684  \\
			& & \textbf{81.6} & & \textbf{73.6} & & \textbf{71.7} & & \textbf{72.7} \\
			\hline
			Veto & 
			25669 & 26656 &
			940   & 799 &
			3270  & 3107 &
			730   & 645 \\
			& & \textbf{89.4} & & \textbf{73.6} & & \textbf{79.2} & & \textbf{75.9} \\
			\hline
			Ordering & 
			25669 & 26656 &
			940   & 799 &
			3082  & 2857 &
			682   & 591 \\
			\textbf{$\nu_{\mu}$CC$-\mu[\%]$} & & \textbf{89.4} & & \textbf{73.6} & & \textbf{81.9} & & \textbf{78.9} \\
		\end{tabular}
	\end{ruledtabular}
	\caption{The selected number of events and signal purities percentage (in bold) in each sample as successive 
	requirements are added for data and MC. The cut in last row refers to the priority order in cases where a muon 
	candidate has been found in two samples.}
	\label{table:reduction_per}
\end{table*}

Table~\ref{table:purity} breaks down each sample in the different reaction channels. In the low angle selections, the 
dominant background is associated with negative pions which are mis-identified as muons in the TPC. In the high angle 
selections, in which there are no TPC segments, positive pions are the dominant background because the charge of the 
track is not reconstructed. Those pions are coming mainly from NC interactions or CC-DIS interactions. For the out of FV 
events, a primary contribution arises from interactions taking place in the borders of FGD1, where the hits closest to the 
interactions are not reconstructed. In the case of interactions in BarrelECal, backgrounds arise when FGD and BarrelECal 
reconstructed segments are not matched. Finally, the contribution from interactions in the P0D is composed primarily by 
neutral particles that scatter inside FGD1.

\begin{table}
	\begin{ruledtabular}
		\begin{tabular}{ c | c | c | c | c  }
			\multicolumn{1}{c|} {} & \multicolumn{1}{c|}{FWD} & \multicolumn{1}{c|}{BWD} & \multicolumn{1}{c|}{HAFWD} & \multicolumn{1}{c}{HABWD} \\
			\hline
			\textbf{$\nu_\mu$CC-$\mu$}    & \textbf{89.4} & \textbf{73.6} & \textbf{81.9} & \textbf{78.9} \\
			\hline
			QE                 & 44.7 & 82.0 & 67.3 & 83.2 \\
			2p2h                 & 7.5  & 5.5  & 7.2  & 5.5  \\
			RES                  & 25.4 & 8.6  & 17.6 & 8.0  \\
			DIS                  & 19.9 & 3.8  & 7.2  & 3.4  \\
			COH                  & 2.5  & 0.0  & 0.7  & 0.0  \\
			\hline
			\textbf{$\nu_\mu$CC-no$\mu$}  & \textbf{2.2} & \textbf{1.1} & \textbf{2.6} & \textbf{1.6} \\
			\hline
			QE                 & 1.8  & 4.5  & 6.3  & 3.0  \\
			2p2h                 & 0.3  & 0.0  & 1.6  & 0.6  \\
			RES                  & 6.3  & 24.6 & 59.1 & 60.8 \\
			DIS                  & 91.4 & 70.3 & 31.7 & 35.6 \\
			COH                  & 0.3  & 0.6  & 1.3  & 0.0  \\
			\hline
			\textbf{no$\nu_\mu$CC} & \textbf{3.3} & \textbf{1.4} & \textbf{3.7} & \textbf{1.9} \\
			\hline
			NC                   & 75.5 & 67.2 & 51.4 & 69.1 \\
			$\bar{\nu}_{\mu}$    & 15.8 & 15.8 & 39.3 & 15.4 \\
			$\nu_{e}, \bar{\nu}_{e}$ & 8.7 & 17.0 & 9.3 & 15.5 \\
			\hline
			\textbf{Out of FV} & \textbf{4.4} & \textbf{21.5} & \textbf{11.3} & \textbf{16.9} \\
			\hline
			$\nu_\mu$CC (in FGD1)  & 12.4 & 16.4 & 33.3 & 34.6 \\
			$\nu_\mu$CC (out FGD1) & 65.2 & 69.2 & 51.7 & 55.9 \\
			NC                     & 17.0 & 11.0 & 11.3 & 7.5  \\
			Other                  & 5.4  & 3.4  & 3.8  & 2.0  \\
			\hline
			\textbf{Sand $\mu$}  & \textbf{0.8}  & \textbf{2.3}  & \textbf{0.4}   & \textbf{0.7} \\
		\end{tabular}
		\caption{Muon candidate composition in NEUT combining the true inclusive reaction type and the true particle 
		type of the muon candidate in bold. The true reaction composition for each topology is shown as plain text.}
		\label{table:purity}
	\end{ruledtabular}
\end{table}

\subsection{Reconstruction efficiencies}

\begin{figure}
	\includegraphics[width=.45\textwidth]{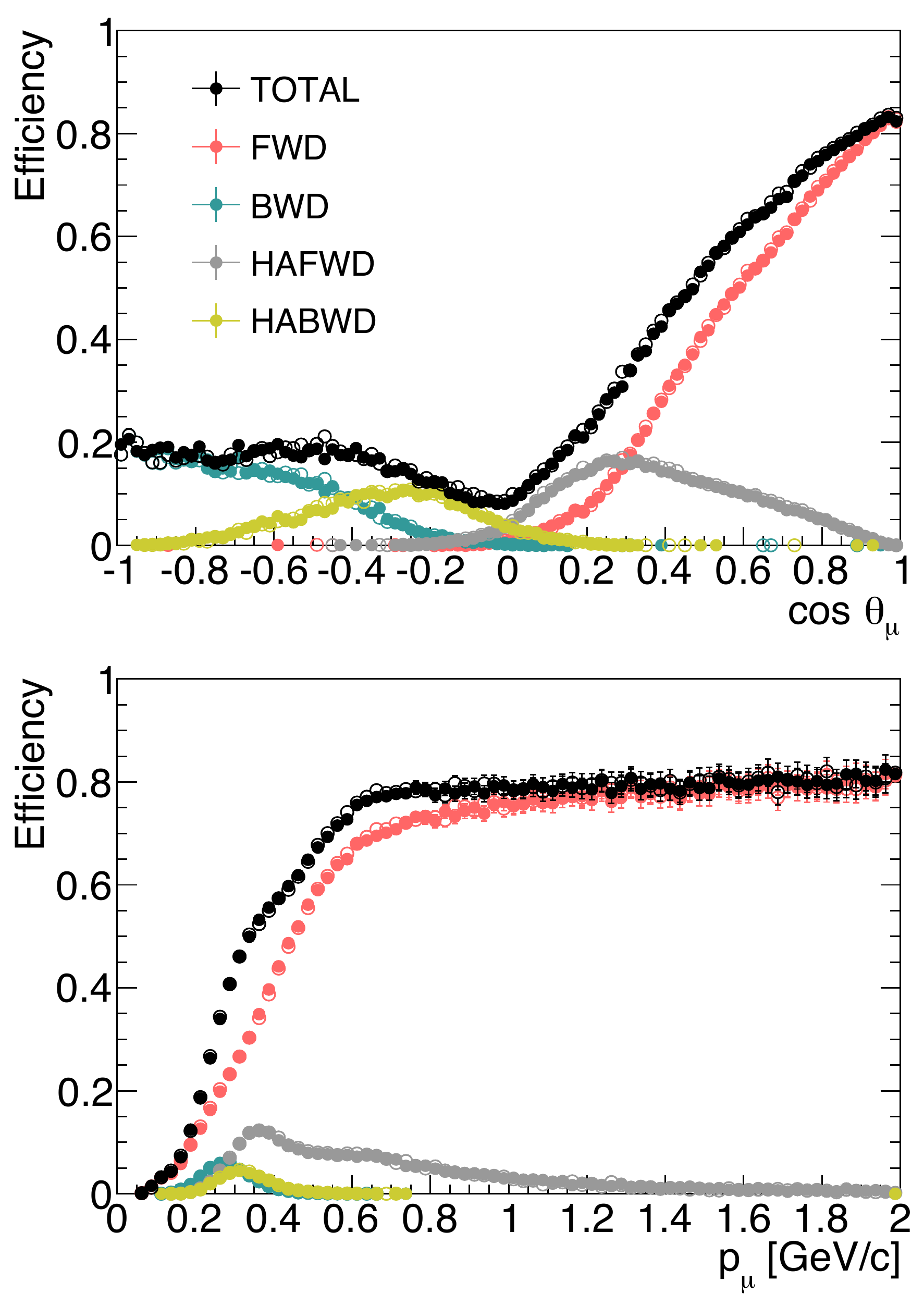}
	\caption{The reconstructed signal efficiency as a function of the momentum and cosine of the emission angle of the true 
	muon using NEUT (full dots) and GENIE (empty dots). The colors indicate contributions from different samples.}
	\label{fig:efficiency}
\end{figure}

The reconstruction efficiency for $\nu_{\mu}$ CC events as a function of the kinematics of the outgoing muon is shown in 
Fig.~\ref{fig:efficiency}. For low momentum muons (below 500 MeV/c) the efficiency drops drastically because such low 
momentum particles are unlikely to exit the FGD and pass the selection criteria. The stopping requirement, necessary to 
determine muon momentum by range and the timing, poses a significant limitation for high angle muons. This is particularly 
true for backward going muons, which occur typically at very low momentum and stop in the passive edges of material 
between subdetectors.  

\begin{figure*}
	\includegraphics[width=1\textwidth]{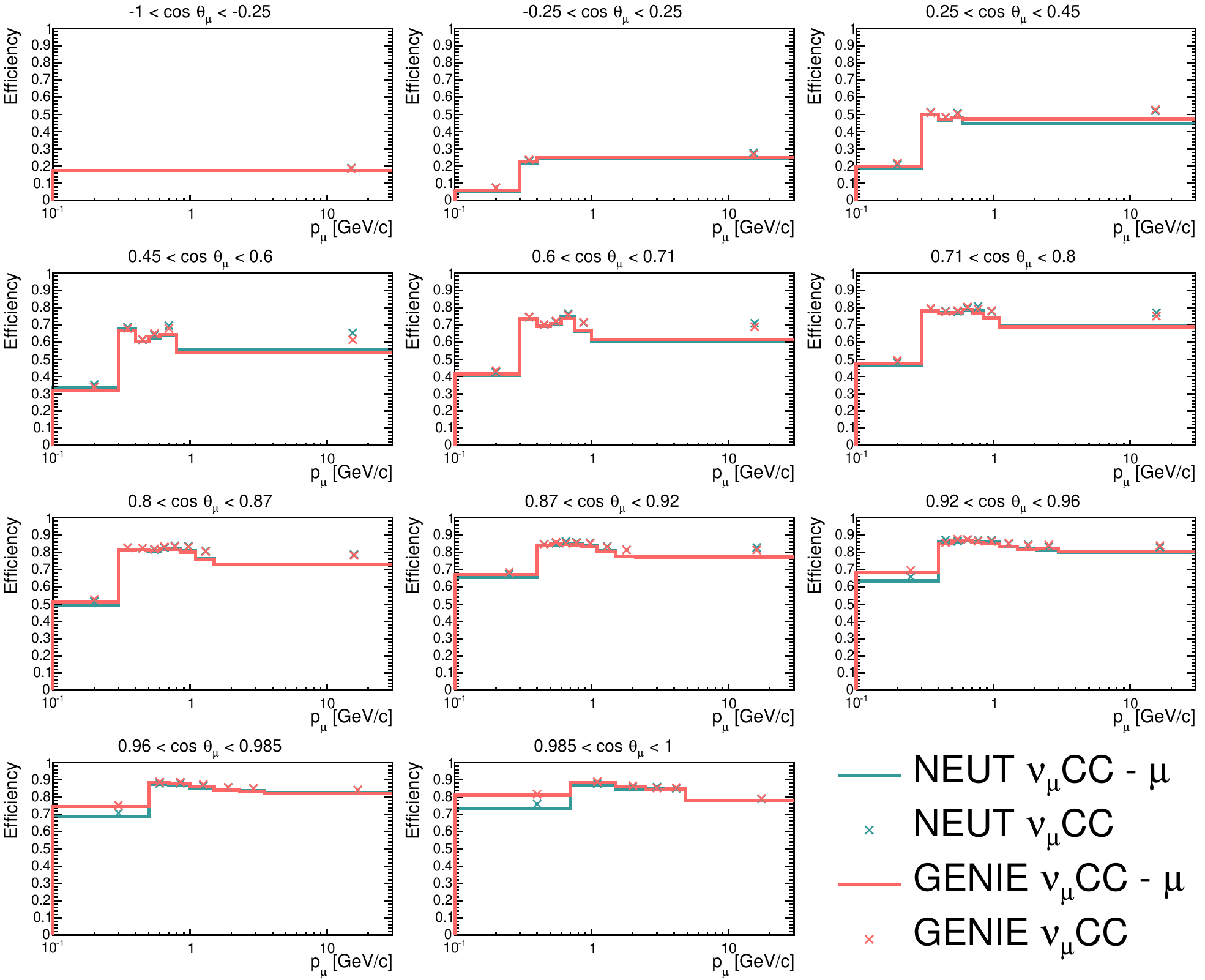}
	\caption{The reconstructed signal efficiency as function of the momentum and cosine of emission angle of the true 
	muon using the same binning as that for the cross-section result (see Table~\ref{table:binning}). Lines represent the 
	efficiencies where the signal is defined as $\nu_{\mu}$ CC events in which the muon candidate is the true muon, so 
	called $\nu_\mu$CC-$\mu$ events. Markers are efficiencies when the muon candidate requirement is not imposed 
	in the sample labeled as $\nu_\mu$CC.}
	\label{fig:eff_mom_all}
\end{figure*}

Fig.~\ref{fig:eff_mom_all} shows the signal reconstruction efficiency using the same binning in $p_\mu$ and 
$\cos \theta_\mu$ as in the cross-section result  (see Table~\ref{table:binning}). The efficiency for high multiplicity events is 
reduced by the fact that $\nu_\mu$ CC events in which the muon candidate is not the true muon (the so called 
$\nu_\mu$CC-no$\mu$ sample) are not included as signal.

The efficiency as calculated in NEUT and GENIE is generally in agreement. However, the predicted efficiency is different for 
low momentum muons going very forward with respect to the neutrino direction. While generators are in principle only used 
to correct for detector effects, this difference highlights how the simulation of final state particles is important even for an 
inclusive selection. In that region of phase space the two generators differ in their predictions for CC deep inelastic and CC 
resonance channels, particularly in the kinematics of the muon and hadrons.

\subsection{Detector systematic uncertainties}

The uncertainties associated with the prediction of each subdetector response (TPCs, FGDs, ECal modules, P0D and SMRD) 
are evaluated using dedicated control samples in the data.  This works since the events in the control samples share many of 
the properties of the events in the $\nu_\mu$ CC selection. 

The tracker systematic uncertainties are divided into four classes: selection efficiency (TPC cluster finding, TPC track finding 
and charge assignment), TPC momentum resolution, TPC PID, and TPC-FGD matching efficiency. They are all assessed as 
in previous analyses from T2K using different control samples of through-going muons \cite{LongOA}.

Uncertainties associated with the ECal modules are computed for the ECal PID, the energy resolution and scale, and the 
efficiency with which ECal objects are reconstructed and matched to TPC tracks. The method to evaluate those errors is 
unchanged with respect to \cite{NuE}, using high purity control samples of muons crossing the TPCs and ECals.

Relative to the previous analysis, this work includes six additional systematic errors. The new errors incorporated in this 
analysis are associated with the ToF; the matching efficiency between TPC-P0D and FGD-ECal(SMRD); the resolution 
of the momentum determined by range; vertex migration; and the neutrino parent direction.

The ToF between FGD1 and FGD2 or BarrelECal or P0D is used to determine if the track starts or ends in the FGD1, and 
infer the charge of the track. The uncertainty is evaluated by comparing the ToF distribution in control samples of tracks 
crossing the relevant subdetectors and starting/stopping in FGD1 for data and MC. The ToF distributions are fit with 
Gaussian distributions for data and simulation. To account for the differences in the means and widths of the distributions 
between data and simulation, corrections are applied to the simulation and the error is set to be equal to the maximum 
bias or resolution correction. The error is not higher than $10\%$ for the Gaussian parameter in any of the distributions.

The TPC-P0D matching efficiency is estimated using a control sample of cosmic muons passing through part of the P0D and 
having a reconstructed segment in TPC1. The efficiency is defined as the ratio between the number of events with a matched 
TPC1-P0D segment and the total number of events in the control sample. This efficiency is evaluated as function of the 
momentum of the track.  The data and MC are less than $5\%$ different when the momentum of the cosmic is higher than 
200 MeV/c.

To compute the FGD-ECal(SMRD) matching efficiency, a control sample is used that contains through-going muons with a 
BarrelECal (SMRD) segment that points to FGD. In order to mimic the kinematics of the muon candidate, it is required that the 
muon stops within the FGD. The matching efficiency is computed from the ratio between the number of events with a matched 
FGD-BarrelECal (or FGD-BarrelECal-SMRD) segment and the total number of events in the relevant control sample. The 
FGD-BarrelECal (FGD-BarrelECal-SMRD) efficiency is found to be $52\%$ ($55\%$) for simulation and $47\%$ ($45\%$) for 
data. A correction is applied to the simulation to account for this and the correction uncertainty is included in the overall detector 
uncertainty.

The momentum by range resolution is studied using particles in a control sample that are fully contained in ND280, stopping 
inside the FGD and BarrelECal (or SMRD), and crossing at least one TPC. The distribution of the difference between the momentum 
determined by curvature using the TPC segment and the momentum by range are compared in data and MC. No bias is 
observed in such distributions but some difference is seen in the width of the distributions; this is used to set the uncertainty. In 
the case of the BarrelECal (SMRD), the systematic uncertainty is around $10\%$ ($30\%$).

The vertex of the interaction is defined as the reconstructed position of the start of the muon candidate inside the FGD. When the 
multiplicity of particles increases, the reconstruction of the vertex becomes more difficult and the vertex position can migrate. 
These migrations have a non negligible impact on the BWD sample event vertices because back-to-back topologies are common 
in that sample. The main effect is on the reconstructed momentum of the muon candidate inside the FGD because it is 
proportional to track length. The difference between the data and simulation for these migrations is difficult to interpret since it is 
sensitive to the modeling of hadrons. An uncertainty of 7 MeV/c (or $\sim$3 FGD layers), which was computed comparing the 
length of the tracks inside the FGD1 for data and MC, is applied to the reconstructed momentum of the muon candidate.

In this analysis, the angle of the outgoing muon is defined with respect to the neutrino direction. The neutrino direction is 
determined from the position of the vertex in FGD1 and the parent hadron decay point of the neutrino in the decay tunnel. The 
mean position of hadron decays in the decay tunnel has an associated uncertainty. This is taken into account by varying the 
mean parent decay point according to the decay distribution in the beam simulation.

The detector systematic uncertainties are propagated in order to check their impact in the rate of reconstructed events in 
$p_\mu$ and $\cos\theta_\mu$. This analysis follows the methodology described in \cite{LongOA}. The expected number of 
events are scaled using a vector of systematic parameters. Then, the uncertainties in each reconstructed bin and their 
correlations are computed using toy experiments in which the systematics are varied simultaneously. 
Table~\ref{table:detector_systematics} shows the full list of detector systematic effects considered and the associated 
uncertainty in each. 

\begin{table}
	\begin{ruledtabular}
		\begin{tabular}{ c | c | c | c | c  }
			\multicolumn{1}{c|} {} & FWD & BWD & HAFWD & HABWD \\
			\hline
			\multicolumn{5}{c} {Efficiency-like} \\
			\hline
			TPC charge ID eff.     & 0.1  & 0.2  & 0  & 0  \\
			TPC cluster eff.       & $<$0.002 & $<$0.002 & $<$0.002 & $<$0.002 \\
			TPC tracking eff.      & 0.8   & 0.4     & 0.05   & 0.02  \\
			ECal tracking eff.     & 0.2    & 0.2    & 4.1    & 4.9  \\
			ECal PID eff.          & 1.3    & 0      & 0.5    & 0.3  \\	
			TPC-FGD match. eff.  & 0.1    & 0.1    & 0.004  & 0.005  \\
			TPC-ECal match. eff. & 0.5    & 0.8    & 0.2    & 0.1  \\
			TPC-P0D  match. eff. & -      & 3.9    & -      & -  \\
			FGD-ECal match. eff. & -      & -      & 4.7    & 6.5  \\
			FGD-SMRD match. eff. & -      & -      & 11.6   & -  \\
			\hline
			\multicolumn{5}{c} {Normalization-like} \\
			\hline
			Pileup                 & 0.2  & -       & 0.2    & 0.2  \\
			Out of fiducial volume & 0.5  & 1.9     & 1.0    & 2.0  \\
			Sand mu                & 0.1  & 0.1     & 0.02   & 0.03  \\
			Pion secondary int     & 0.3  & 0.3     & 0.4    & 0.3  \\
			Proton secondary int   & 0.01 & 0.001   & 0.2    & 0.01  \\
			\hline
			\multicolumn{5}{c} {Observable variation} \\
			\hline
			TPC Field Distortions           & 0.007 & 0.008  & 0.001  & 0.004  \\
			TPC momentum scale              & 0.007 & 0      & 0.004  & 0.01  \\
			TPC momentum res.         & 0.02  & 0.015  & 0.01   & 0.01  \\
			Vertex migration  & 0.003 & 0      & 0.01   & 0.01  \\
			TPC PID                         & 0.2   & 0.4    & 0.02   & 0.02  \\
			Momentum range res.    & -     & -      & 0.1    & 0.1  \\
			ECal energy resolution          & -     & -      & 0.1    & 0.2  \\
			ECal energy scale               & -     & -      & 0.8    & 1.5  \\
			Time of flight                  & 0.1   & 2.6    & 2.4    & 7.3  \\
			$\nu$ direction                 & 0     & 0      & 0      & 0  \\
			\hline
			Total                 & 1.8     & 5.9      & 14.3      & 14.3  \\
		\end{tabular}
		\caption{A summary of the fractional systematic uncertainty (in percentage) associated with the detector response. 
		The first column lists all the sources taken into account and the other columns show the error size on the predicted 
		events in each sample.}
		\label{table:detector_systematics}
	\end{ruledtabular}
\end{table}

The uncertainty associated to the matching among FGD, ECal and SMRD subdetectors is dominant in both the HAFWD and 
HABWD selections. The reason is that the misalignment between both subdetectors has not been properly corrected in data, 
leading to discrepancies in the matching efficiency for segments contained in those subdetectors. In the case of the BWD 
sample, the matching between the TPC and P0D subdetectors and the ToF resolution dominates. Meanwhile, in the FWD 
selection the uncertainty associated with the particle identification in the BarrelECal and DsECal dominates.

\section{CROSS-SECTION ANALYSIS}
\label{sec:analysis}
The following section describes the procedure to unfold the measured muon kinematic distributions and to propagate 
uncertainties in the cross-section measurement. After this, the flux-integrated, double-differential cross section results for 
$\nu_\mu$ CC interactions are presented.

\subsection{Methodology}
\label{sec:methodology}

The flux-integrated, double-differential cross section is expressed as

\begin{equation}
\label{eq:xsec}
\frac{\text{d}\sigma_{f.i.}}{\text{d}p_\mu \text{d} \cos \theta_\mu} = \frac{S_{ij}^{\nu_{\mu}\text{CC}-\mu}}{\epsilon_{ij}^{\nu_{\mu}\text{CC}-\mu} \Phi \text{N}_{\text{FV}} \Delta p_{\mu,i} \Delta \cos \theta_{\mu,j}}
\end{equation}

where $S_{ij}^{\nu_{\mu}\text{CC}-\mu}$ is the number of signal events with momentum and angle bins $i$ and $j$, 
respectively. $\epsilon_{ij}^{\nu_{\mu}\text{CC}-\mu}$ is the signal reconstruction efficiency with momentum bin $i$ and angle 
bin $j$. $\Delta p_\mu$ and $\Delta \cos \theta_\mu$ represent the bin widths. Finally, the normalization factors are the total 
integrated flux and the number of target nucleons in the FV. 

The number of nucleons is computed using the areal density of the different elements composing the FV 
($\text{N}_{\text{FV}} = (5.93 \pm 0.04)\times 10^{29}$) \cite{FGD}. The integrated muon neutrino flux is 
$\Phi = (1.107 \pm 0.097)\times 10^{13}$ cm$^{-2}$.

The reconstructed momentum and cosine of emission angle of the muon candidate are not an exact representation of the true 
initial muon kinematics. Therefore, an unfolding method is used to remove the detector effects in the measurement. In this 
analysis, we unfold the muon kinematic quantities using a binned likelihood fit as in \cite{CC0pi}. We vary the true spectrum of the 
simulation (so called prior) and propagate its effect to the rate of events in each reconstructed bin. Then, the predicted rate is 
compared with the values from data. The variation of the true spectrum is performed scaling up or down the rate of signal 
events simultaneously in the four signal and two control regions for each true bin. Two of the parameters associated to the 
background modeling (the normalizations of the neutral current cross section and pion final state interactions) are included in 
the fit as nuisance parameters.

This unfolding method is unregularised, which leads to strong anticorrelations between neighboring bins if the binning is not 
properly defined. The different samples described in Sec.~\ref{sec:samples} are well separated in the angular phase space. In 
fact, the detector response is different for the selected events in each sample. Thus,  the angular binning is chosen (i.e. 
$\cos \theta_\mu$) to separate the contribution from each sample as much as possible. The momentum binning reflects the resolution 
of the detector and is chosen to maintain sufficient statistics in each bin. Table~\ref{table:binning} shows the binning used in this 
analysis for the chosen muon kinematic variables. 

\begin{table}
\begin{ruledtabular}
\begin{tabular}{ r | l }
$\cos \theta_{\mu}$ & $p_{\mu}$ [GeV/c] \\
\hline
-1, -0.25   & 0, 30 \\
-0.25, 0.25 & 0, 0.3, 0.4, 30 \\
0.25, 0.45  & 0, 0.3, 0.4, 0.5, 0.6, 30 \\
0.45, 0.6   & 0, 0.3, 0.4, 0.5, 0.6, 0.8, 30 \\
0.6, 0.71   & 0, 0.3, 0.4, 0.5, 0.6, 0.75, 1, 30 \\
0.71, 0.8   & 0, 0.3, 0.4, 0.5, 0.6, 0.7, 0.85, 1.1, 30 \\
0.8, 0.87   & 0, 0.3, 0.4, 0.5, 0.6, 0.7, 0.85, 1.1, 1.5, 30 \\
0.87, 0.92  & 0, 0.4, 0.5, 0.6, 0.7, 0.85, 1.1, 1.5, 2.1, 30 \\
0.92, 0.96  & 0, 0.4, 0.5, 0.6, 0.7, 0.85, 1.1, 1.5, 2.1, 3, 30 \\
0.96, 0.985 & 0, 0.5, 0.7, 1, 1.5, 2.3, 3.5, 30 \\
0.985, 1    & 0, 0.7, 1.5, 2.5, 3.5, 4.8, 30 \\
\end{tabular}
\caption{Binning used for $\cos \theta_{\mu}$ and $p_{\mu}$ distributions in both reconstructed and true phase space.}
\label{table:binning}
\end{ruledtabular}
\end{table}

\subsection{Error propagation}
\label{sec:errors}

Analytical computation for most of the uncertainties in this analysis is not possible. So toy experiments are used to study 
their impact and determine errors. In the toy experiments, some aspect of the simulated or real data is changed depending 
on the source of uncertainty as described below.

To evaluate the uncertainty due to data statistics, toy experiments are produced applying a Poisson fluctuation to the 
number of reconstructed events in the data for each bin and sample. For each toy, the fluctuated data are unfolded using 
as prior the nominal MC and the cross section is computed using Eq.~\ref{eq:xsec}. The statistical error in each bin is 
taken as width of the cross section distribution for many toys.

The methodology used to estimate systematic uncertainties involves reweighting the MC prediction for each toy experiment. 
Parameters associated to each systematic error are thrown according to a Gaussian distribution around the nominal value, 
following the prior errors and taking into account correlations. Then, for each toy, the data is unfolded using as prior the 
reweighted MC. In addition, $\Phi$, $\text{N}_{\text{FV}}$ and $\epsilon_{ij}^{\nu_{\mu}\text{CC}-\mu}$ are also weighted 
using the thrown value of the parameters. Finally, the cross section is computed using Eq.~\ref{eq:xsec} for each toy. The 
uncertainty in each bin is taken as width of the cross section distribution for many toys.

\begin{figure*}
\includegraphics[width=1\textwidth]{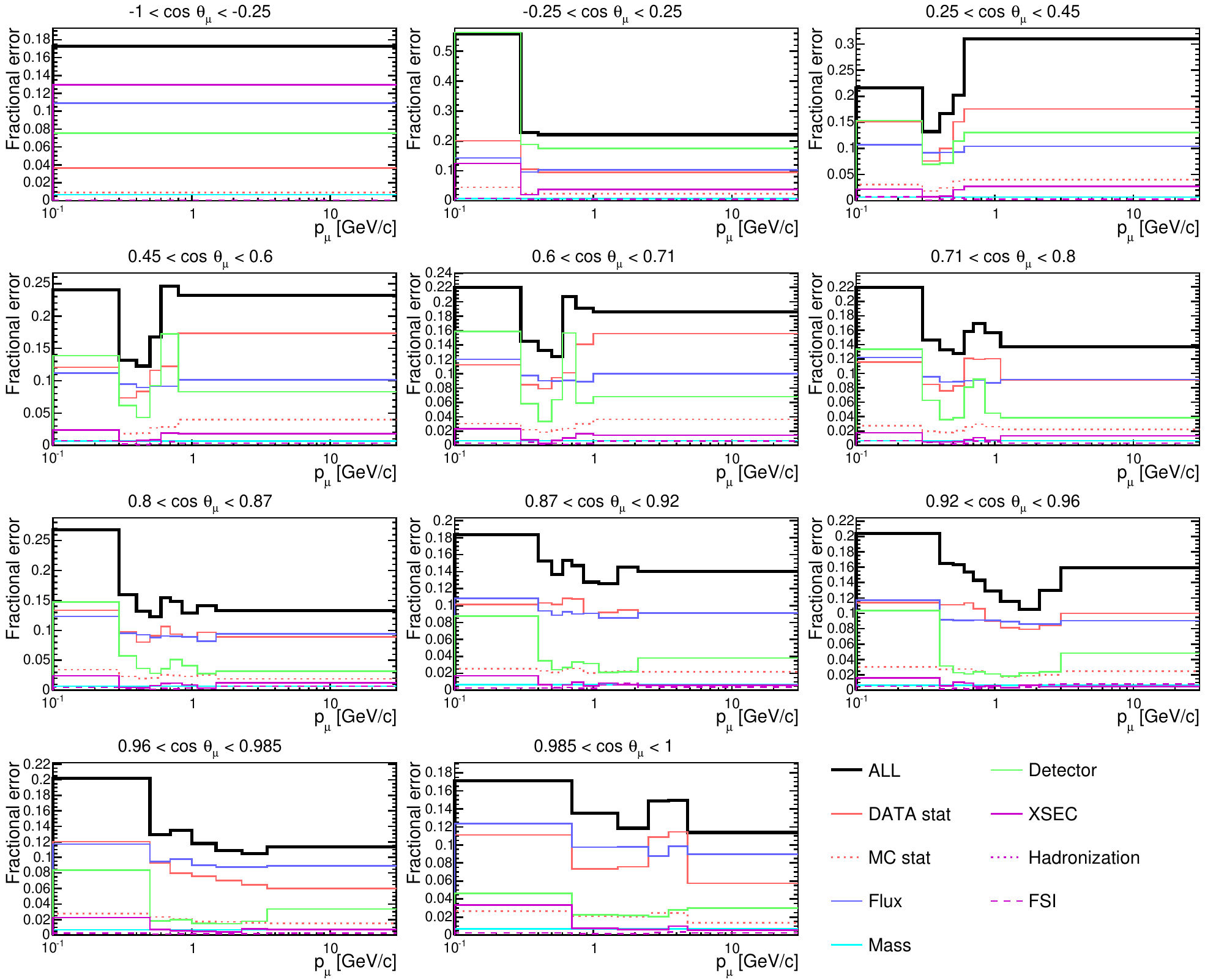}
\caption{The fractional error from each source of uncertainty on the flux-integrated, double-differential cross section. The total 
error is computed varying simultaneously both statistical and systematic uncertainties.}
\label{fig:xsec_error}
\end{figure*}

Fig.~\ref{fig:xsec_error} shows a comparison of the fractional error associated to each source of uncertainty using 1500 toy 
models. Throughout most of the phase space, the dominant systematic uncertainty is the flux. In the backward region, the 
neutrino interaction modeling dominates, with the largest contribution coming from the uncertainty assigned to the $M_A$ 
parameter. The detector systematic becomes relevant in the high angle region ($-0.25<\cos \theta_{\mu}<0.25$) due to the 
large uncertainties in FGD-ECal(SMRD) matching efficiencies, and at very low momentum where the out of FV contribution 
is more pronounced. The statistical uncertainty is dominant in the high momentum region where the number of reconstructed 
events is lower (except at low angles in the forward direction). 

It is interesting to note that the systematic uncertainties associated with the signal and background modeling give a relatively 
unimportant contribution to the overall inclusive cross section uncertainty because of the high purity and efficiency for the 
signal sample. The systematic uncertainties associated with the modeling of neutral-current interactions and pion final state 
interactions are reduced by a factor of 2 thanks to the use of the control samples.

\section{RESULTS AND CONCLUSIONS}
\label{sec:results}
The flux-integrated total cross section is computed by integrating both the number of signal events and the signal efficiency 
over the muon phase space.  

\begin{eqnarray*}
	\sigma_{\text{DATA FIT W/ NEUT}} = \left( 6.950 \pm 0.049[\text{stat}] \pm 0.123[\text{syst}] \right. \\
								\left. \pm 0.608[\text{flux}] \right) \times 10^{-39} \text{cm}^{2}\text{nucleon}^{-1} \\
	\sigma_{\text{DATA FIT W/ GENIE}} = \left( 6.850 \pm 0.048[\text{stat}] \pm 0.121[\text{syst}] \right. \\
								\left. \pm 0.599[\text{flux}] \right) \times 10^{-39} \text{cm}^{2}\text{nucleon}^{-1} \\
\end{eqnarray*}

This is compatible with predictions from the two event generators: 
$\sigma_{\text{NEUT}} = 7.108 \times 10^{-39} \text{cm}^{2}\text{nucleon}^{-1}$ and $\sigma_{\text{GENIE}} = 6.564 \times 10^{-39} \text{cm}^{2}\text{nucleon}^{-1}$. 
It is known that the detector performance varies substantially as a function of the momentum and angle of the outgoing muon. 
Therefore, the extracted value using the total cross section must be interpreted cautiously. This result shows good agreement 
with the one obtained in \cite{T2K-CCincl}.

\begin{figure*}
\includegraphics[width=1\textwidth]{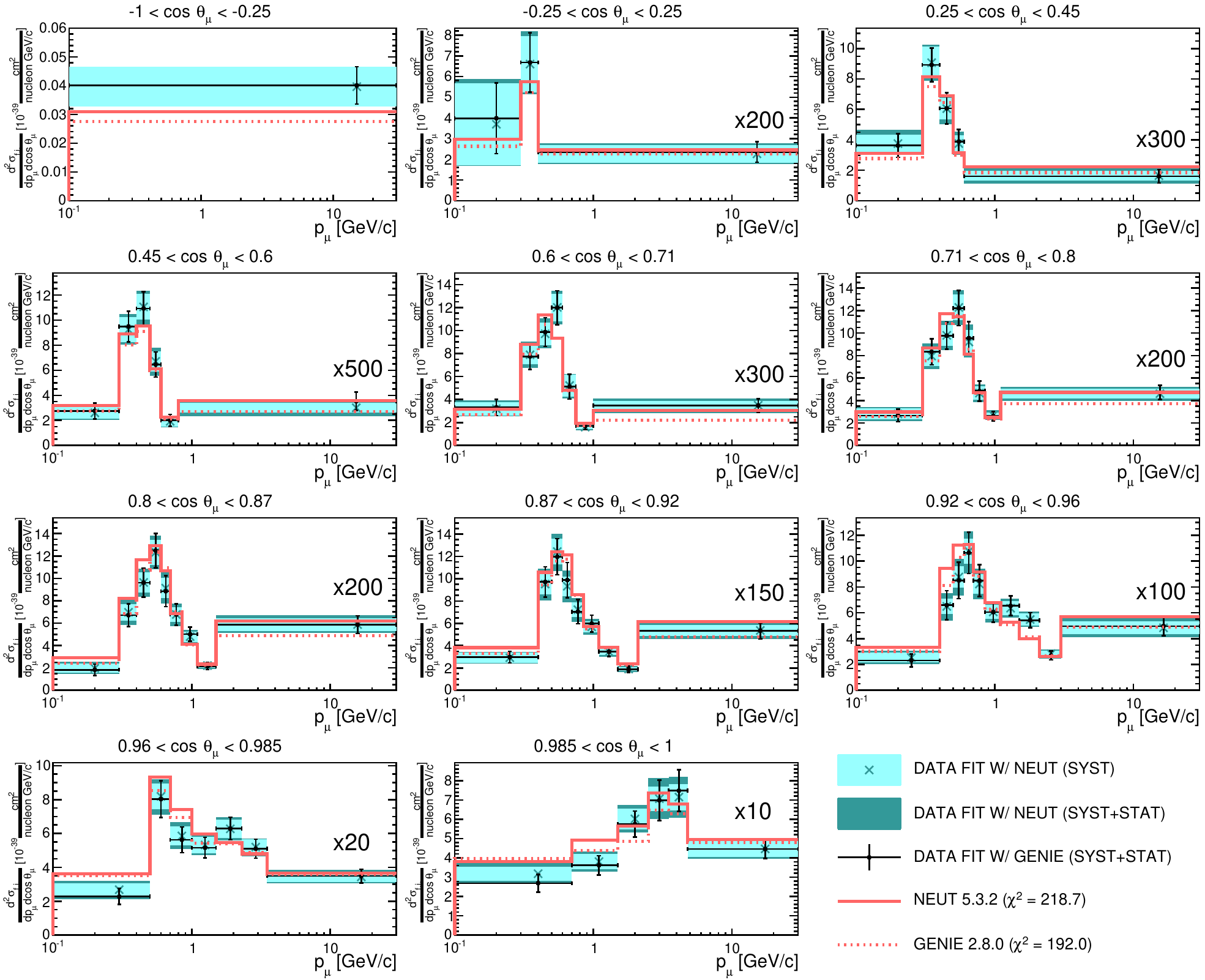}
\caption{The flux-integrated, double-differential cross section per nucleon for NEUT (continuous red line), for GENIE (dashed 
red line), and the unfolded-data result using as prior either NEUT or GENIE. The bin of highest momentum is scaled by the 
factor shown in each plot to make it visible. $\chi^2$ values are computed with unfolded-data result using as prior NEUT.}
\label{fig:xsec}
\end{figure*}

The flux-integrated, double-differential cross section is computed as function of the outgoing muon kinematics using the 
methodology described in Sec.~\ref{sec:methodology} and Sec.~\ref{sec:errors} using two independent MC generators 
detailed in Sec.~\ref{sec:generators}. Fig.~\ref{fig:xsec} shows the results for the unfolded data as well as the NEUT and 
GENIE predictions. A small disagreement is observed in the low momentum and very forward regions when using different 
event generators as prior. This bias is not due to unfolding but due to the different efficiency corrections in that region of the 
phase space for NEUT and GENIE as shown in Fig.~\ref{fig:eff_mom_all}. The muon neutrino flux used in this analysis and the 
measured cross section values, errors and correlation matrix can be found in \cite{datarelease}.

This result is compared to the NEUT and GENIE predictions, showing in both cases high $\chi^2$ values with respect to the total 
number of bins, 71. In the new regions of phase space (high angle and backward-going muons) there is good agreement but 
uncertainties are still large. For forward-going muons the binning is finer and interesting structures are observed. 

\section*{ACKNOWLEDGEMENTS}
We thank the J-PARC staff for superb accelerator performance. We thank the CERN NA61/SHINE Collaboration for providing 
valuable particle production data. We acknowledge the support of MEXT, Japan; NSERC (Grant No. SAPPJ-2014-00031), 
NRC and CFI, Canada; CEA and CNRS/IN2P3, France; DFG, Germany; INFN, Italy; National Science Centre (NCN) and 
Ministry of Science and Higher Education, Poland; RSF, RFBR, and MES, Russia; MINECO and ERDF funds, Spain; SNSF 
and SERI, Switzerland; STFC, UK; and DOE, USA. We also thank CERN for the UA1/NOMAD magnet, DESY for the HERA-B 
magnet mover system, NII for SINET4, the WestGrid and SciNet consortia in Compute Canada, and GridPP in the United 
Kingdom. In addition, participation of individual researchers and institutions has been further supported by funds from ERC 
(FP7), H2020 Grant No. RISE-GA644294-JENNIFER, EU; JSPS, Japan; Royal Society, UK; the Alfred P. Sloan Foundation 
and the DOE Early Career program, USA.

\bibliographystyle{apsrev4-1}
\bibliography{prd}

\end{document}